\documentclass[lettersize,journal]{IEEEtran}

\def\BibTeX{{\rm B\kern-.05em{\sc i\kern-.025em b}\kern-.08em
    T\kern-.1667em\lower.7ex\hbox{E}\kern-.125emX}}
\usepackage{balance}

\usepackage{amsmath,amssymb,amsfonts}
\usepackage{algorithmic}
\usepackage{algorithm}
\usepackage{graphicx}
\usepackage{textcomp}
\usepackage[cmyk]{xcolor}
\usepackage{color}
\usepackage{booktabs}
\usepackage{verbatim}
\usepackage{comment}
\usepackage{multirow}
\usepackage{multicol} 
\usepackage{bbding}

\usepackage[caption=false,font=normalsize,labelfont=sf,textfont=sf]{subfig}

\begin{document}

\title{Magnifier: Detecting Network Access via Lightweight Traffic-based Fingerprints}

\author{Wenhao Li,~\IEEEmembership{Member,~IEEE,}
Qiang Wang,~\IEEEmembership{}
Huaifeng Bao,~\IEEEmembership{Student Member,~IEEE,}
	Xiao-Yu Zhang,~\IEEEmembership{Senior Member,~IEEE,}
	Lingyun Ying,~\IEEEmembership{}
	Zhaoxuan Li,~\IEEEmembership{Student Member,~IEEE}
}

\markboth{Magnifier: Detecting Network Access via Lightweight Traffic-based Fingerprints}
{}

\maketitle

\begin{abstract}

Network access detection plays a crucial role in global network management, enabling efficient network monitoring and topology measurement by identifying unauthorized network access and gathering detailed information about mobile devices. Existing methods for endpoint-based detection primarily rely on deploying monitoring software to recognize network connections. However, the challenges associated with developing and maintaining such systems have limited their universality and coverage in practical deployments, especially given the cost implications of covering a wide array of devices with heterogeneous operating systems. To tackle the issues, we propose Magnifier for mobile device network access detection that, for the first time, passively infers access patterns from backbone traffic at the gateway level. Magnifier's foundation is the creation of device-specific access patterns using the innovative Domain Name Forest (dnForest) fingerprints. We then employ a two-stage distillation algorithm to fine-tune the weights of individual Domain Name Trees (dnTree) within each dnForest, emphasizing the unique device fingerprints. With these meticulously crafted fingerprints, Magnifier efficiently infers network access from backbone traffic using a lightweight fingerprint matching algorithm. Our experimental results, conducted in real-world scenarios, demonstrate that Magnifier exhibits exceptional universality and coverage in both initial and repetitive network access detection in real-time. To facilitate further research, we have thoughtfully curated the NetCess2023 dataset, comprising network access data from 26 different models across 7 brands, covering the majority of mainstream mobile devices. We have also made both the Magnifier prototype and the NetCess2023 dataset publicly available\footnote{https://github.com/SecTeamPolaris/Magnifier}.

\end{abstract}


\section{Introduction}

\IEEEPARstart{N}{etwork} management is the art of grasping information as much as possible in the authorized networking domain~\cite{cao2014survey,wright2006inferring,prograph}. Within the realm of corporate networks, particularly those bound by stringent confidentiality protocols, a high degree of network monitoring is imperative. In these complex network configurations, unauthorized network access by devices is unequivocally prohibited due to the inherent security risks, including an elevated potential for breaches of privacy~\cite{maithili2018analyzing,jana2008fast}. Personal devices lacking official authorization, including mobile phones and personal computers, have the capability to infiltrate internal networks by leveraging wireless hotspots facilitated by authorized devices and routers. This method of access remains surreptitious and elusive to network administrators.

\begin{figure}[t]
	\centering

	\includegraphics[width=8cm]{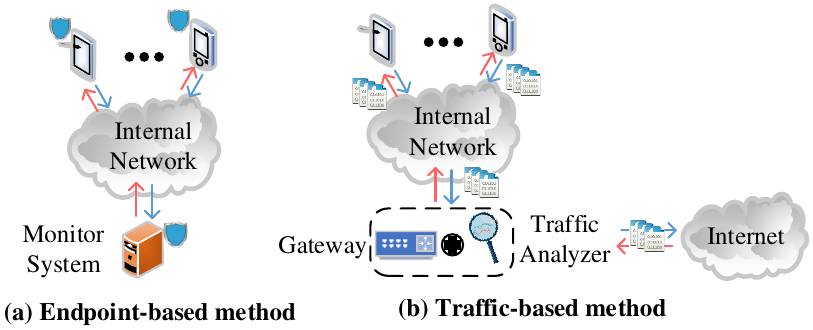}
	\caption{Two types of methods for network access detection.}
	\label{sc_fig}

\end{figure}

Conventionally, network operators have employed active methods to discern network access behaviors, utilizing software that can be readily controlled and deployed on authorized devices, as shwon in Figure~\ref{sc_fig}.(a). These methods have undergone extensive exploration both in industry and academia~\cite{hassan2020tactical,sjarif2019endpoint}. A prominent example is the implementation of Endpoint Detection and Response (EDR) systems, renowned for their capacity to identify access behaviors as soon as a device with monitoring software connects to the network. While client/server-based monitoring systems of this nature excel at the precise detection of network access, the underestimation of the costs associated with the development and maintenance of such systems presents practical deployment challenges for these approaches.


\begin{itemize}

\item \textbf{Poor universality} arises from the complexities of developing monitoring software for diverse platforms and operating systems (OS). The associated maintenance, including debugging and version updates, adds further cost implications to these monitoring solutions. 

\item \textbf{Coverage is limited} due to the diversity of devices, making it difficult to deploy monitoring software on all potentially accessing devices, particularly those under bring-your-own-device (BYOD) policies. Privacy regulations can further impede software deployment, hindering the detection of mobile devices with intermittent network access.

\end{itemize}


With the growth of network traffic analysis, passive inference of access behavior from network traffic offers a path to bypass the constraints imposed on endpoint-based methods~\cite{li2022robust,wang2020symtcp,elmasry2020evolving, flowprint,velan2015survey,mampf}. Traffic-based analyzers enhance accessibility for network management operators who lack direct control over devices, particularly in BYOD policy environments. As shown in Figure~\ref{sc_fig}.(b), deployed at the authorized network's gateway, these analyzers can effectively detect network access and discern device models with improved universality and coverage in comparison.

\begin{itemize}
\item \textbf{Universality. }Despite the OS and platform heterogeneity, traffic-based methods can detect network access without the need for deploying and maintaining monitoring software, provided that the device traffic can be captured. 

\item \textbf{High Coverage. }A singular deployment on the gateway can effectively monitor the majority of network access within the authorized network, provided devices establish connections with servers outside the domain.

\end{itemize}

Despite the success of network traffic analysis techniques, effective methods for detecting device network access are limited due to two primary challenges. First, subnets created by internal routers and access points are hidden from the gateway view, making static destination-based methods impractical~\cite{wang2023dialectical}. Second, device network access is temporally unpredictable, and access bursts can easily be obscured by backbone traffic. Learning-based methods, which rely on complex feature engineering and decision logic, are ill-suited for online detection systems that require rapid response and resource efficiency.

\textbf{Upon the inclusion of a mobile device into the network, a distinctive traffic pattern becomes discernible from the gateway's perspective.} Based on this observation, this paper proposes Magnifier, designed for the detection of network access behaviors in various devices, with a particular emphasis on mobile devices. To overcome the first challenge, Magnifier employs fingerprinting for each mobile device, capturing network traffic patterns during their connection to the authorized network. Notably, these constructed fingerprints are independent of IP addresses, making Magnifier resilient to dynamically changing addresses. Addressing the second challenge, Magnifier presents an algorithm for constructing lightweight fingerprints of device network access. Furthermore, a lightweight fingerprint matching algorithm is introduced to facilitate real-time detection of mobile device network access, offering practical deployment within online network management systems.

Generally, our contributions can be briefly summarized as follows:
\begin{enumerate}

\item To the best of our knowledge, Magnifier represents one of the pioneering efforts in detecting device network access, with a particular focus on mobile devices, thereby enhancing network management capabilities.

\item Empowered by meticulously crafted fingerprint-based algorithms, Magnifier achieves real-time detection of mobile device network access, offering lightweight deployment and rapid responsiveness. Empirical experiments in real-world scenarios affirm the efficacy and efficiency of Magnifier.


\item We meticulously compiled a dataset, namely NetCess2023, capturing network access traffic from 26 mobile device models, covering the mainstream mobile phone landscape. This dataset is now publicly available to encourage further research and advance real-world network applications\footnote{https://github.com/SecTeamPolaris/Magnifier}.
 
\end{enumerate}

\textbf{Problem Scope. }In this paper, we concentrate on detecting network access behaviors of mobile devices through internal wireless access points, achieved by analyzing backbone traffic and inferring detailed device information at the gateway. 

The rest of the paper is organized as follows. In Section~\ref{sec_related}, we briefly summarize the existing methods in network access detection and network traffic analysis. In Section~\ref{sec_pre}, we preliminarily introduce the scenarios of this paper. In Section~\ref{sec_method}, we describe Magnifier in detail. In Section~\ref{sec_expset}, we introduce our dataset in detail. In Section~\ref{sec_expres}, Magnifier is evaluated under several real-world scenarios, compared with the existing methods. We conclude our work in Section~\ref{sec_conclusion}.

\section{Background and Related Work}\label{sec_related}

\subsection{Network Access Detection}

Deploying monitoring software on devices prior to network access allows for precise network access detection. Consequently, many enterprise network operators employ these monitoring systems. Nevertheless, as previously noted, developing and maintaining these systems is resource-intensive, given the diversity of devices in networks.

Radio frequency-based methods, conversely, identify hotspot access points by intercepting WiFi signals using specialized devices~\cite{xia2004detecting,rayanchu2011airshark,abdelnasser2015wigest}. However, they exclusively detect network access via wireless access points. Furthermore, deploying such radio detectors entails significant financial investment, often resulting in limited coverage.

\subsection{Network Traffic Analysis}
Network traffic analysis has been one of the most heated fields in network management~\cite{zhang2014robust,majeed2020cross,lstm,ren2021tree,li2023prism}. Existing methods classify the network traffic from packets or flows~\cite{malekghaini2023deep,bovenzi2023few,dai2023glads}. A flow is a set of packets in the form of five-tuple, \textit{\{Destination IP, Source IP, Destination Port, Source Port, Protocol\}}. The prosperity of artificial intelligence empowers the identification of encrypted network traffic~\cite{datanet,acnn,acgan,RBRN,fu2023detecting,lin2022bert}, particularly in application identification~\cite{oh2023appsniffer,zhang2023tfe} and website fingerprinting~\cite{huang2023efficient,de2019poster,de2020trafficsliver}. 


FS-Net~\cite{fsnet} employs an AutoEncoder with Gated Recurrent Unit (GRU) to analyze network flows, achieving a 0.99 TPR in correlating flows from 18 popular applications. MBTree~\cite{mbtree} uses packet size sequences to detect malicious network traffic with a tree-based approach. FC-Net~\cite{fcnet} employs meta-learning on grayscale images converted from byte sequences to analyze network traffic. FlowPrint~\cite{flowprint} classifies Android network traffic using statistical features and certificates.

Although only a few analyzers claim to detect device network access, we draw inspiration from related ideas in network management~\cite{yuan2023mcre,10262062}. Our main emphasis in this paper is the comparison of Magnifier with fine-tuned state-of-the-art network traffic analyzers.

\begin{figure*}[t]
	\centering

	\includegraphics[width=\linewidth ]{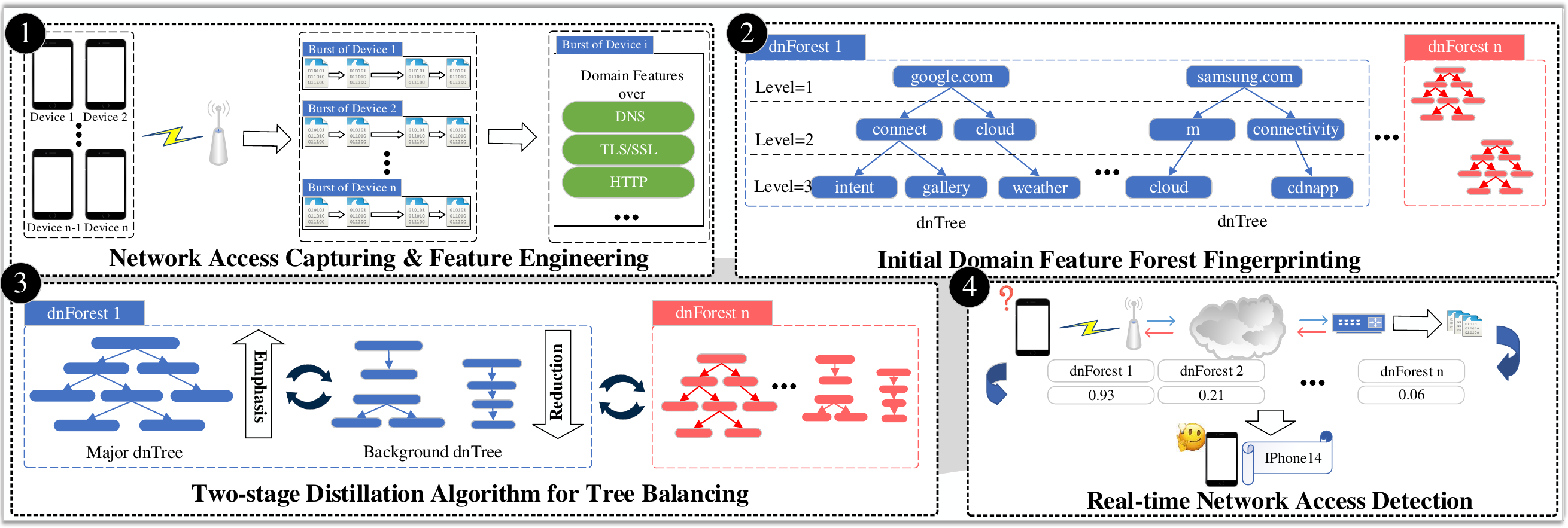}
	\caption{The framework of Magnifier. }
	\label{mainstructure}

\end{figure*}

\section{Preliminaries}\label{sec_pre}

\subsection{Motivation}

Network access by unauthorized devices can pose security threats to enterprise networks, particularly those with stringent confidentiality requirements. Given the constraints of endpoint-based methods, we advocate for detecting network access and identifying device models through gateway-based network traffic analysis.Whenever a device connects to the network, communication traffic becomes visible on the gateway as long as the device establishes a connection with a remote server. To address this, a detector must perform comprehensive traffic monitoring at the gateway with fine-grained detection capabilities and rapid response times, since network access bursts are unpredictable, inconspicuous, and transient.

\subsection{Problem Definition}
In this paper, we mainly focus on detecting network access established by mobile devices. Assume that $T_D$ is the traffic generated by device $y$ when it joins the network. $T_G$ is the traffic captured on the gateway of the authorized network. Note that $T_G$ contains both $T_D$ and irrelevant traffic (i.e., $T_D \subset T_G$). Our goal is to detect the corresponding network accessing in $T_G$ (i.e., $f(T_G, \theta) = y$), where $f(\cdot)$ is the detector with parameters $\theta$. Especially, if multiple devices join the network during the same detection window (e.g., $\{T_D^a, T_D^b, T_D^c\} \subset T_G$ where $T_D^a$, $T_D^b$ and $T_D^c$ are generated by device $a$, $b$ and $c$ respectively), the detector should distinguish all possible network access (i.e., $f(T_G, \theta) = \{y^a, y^b, y^c\}$).

\section{Magnifier} \label{sec_method}

In this section, we outline the methodology for detecting and classifying mobile device network access using Magnifier. As shown in Figure~\ref{mainstructure}, Magnifier initially creates fingerprints for each mobile device's network access. This process involves extracting domain features from various traffic burst protocols and constructing a domain name forest (dnForest) for each device. To refine these fingerprints, we introduce a two-stage distillation algorithm. Initially, we adjust the domain name trees (dnTree) within each dnForest by emphasizing the primary dnTree and reducing the significance of less relevant dnTrees, enhancing intra-class scatter. Subsequently, we reconfigure the inter-class scatter using a tree-based TF-IDF algorithm. With these enhanced fingerprints, Magnifier is deployed on the gateway to monitor real-time traffic, facilitating the detection of network access behavior and mobile device model classification.

Briefly, the overall workflow of Magnifier is as follows:
\begin{itemize}
\item Magnifier constructs a dnForest for each device by fingerprinting the domain features of the traffic burst generated during network access.

\item Magnifier employs a two-stage distillation algorithm to highlight proprietary fingerprints while reducing the importance of background ones.



\item With balanced fingerprints, Magnifier achieves real-time detection of network access on mobile devices using a lightweight matching algorithm.

\end{itemize}

\subsection{Network Access Fingerprinting}
Before performing online detection of network access efficiently, we firstly model domain features of mobile devices with lightweight fingerprints.

\subsubsection{Feature Extraction} \label{sec_features}

Personal mobile devices are primarily divided into two dominant platforms: Android and macOS. Both of these platforms are derived from the Linux/Unix system. To be more specific, manufacturers develop their unique user interfaces (UI), comprising a series of proprietary applications or components. For example, an iPhone with macOS comes pre-installed with the Apple Store and iCloud, while a Huawei P50 integrates exclusive components like Huawei Weather or Huawei Pay into HarmonyOS.

Despite this diversity, every time these devices connect to the network, their integrated components or applications attempt to establish connections with servers relevant to their respective brands. These connections are essential for exchanging up-to-date information or accessing remote services. Theoretically, connections with different brand-relevant servers can result in distinct network traffic patterns. Based on this theory, we create fingerprints for these connections to model the network access of various mobile devices.

Network access traffic is typically encrypted using TLS/SSL or private protocols over HTTP. Traditional methods for encrypted traffic classification (ETC) rely on side-channel features like statistical and temporal attributes. However, these features are susceptible to instability, causing performance issues in out-of-distribution scenarios~\cite{prograph}. In contrast, domain features have proven to be more reliable. In practical deployments, mobile devices consistently request a series of services upon network access, whether directly through WiFi or transitioning from 4G or 5G. We can extract domain features from this network traffic, even in the presence of encryption. 

In this paper, our primary focus is on extracting domain features from commonly used protocols within network access traffic bursts. Typically, these domain features are derived from three typical protocols.

\textbf{Domain Features over DNS. }In addition to network access, devices send DNS requests to resolve the IP addresses of target servers before establishing connections with remote services. While the subsequent communication is encrypted, DNS requests preceding the encrypted traffic are observable, allowing us to extract domain features. This involves extracting domain names from the queries and responses within the DNS traffic accompanying the network access bursts of each mobile device.

\textbf{Domain Features over TLS/SSL. }When a device joins a network, background applications and components are activated to establish communication with brand-specific remote servers. Although the payload remains entirely concealed by TLS/SSL encryption, the handshakes are transmitted in plaintext. In these handshakes, we can distinguish server information from the encryption certificates. We extract server names from the standard TLS/SSL encrypted traffic that accompanies the network access burst of each mobile device.

\textbf{Domain Features over HTTP. }In addition to standard TLS/SSL encryption, customized encryption based on proprietary algorithms is prevalent in network access traffic. Typically, the encrypted payload is encapsulated within the Hypertext Transfer Protocol (HTTP), owing to its ease of maintenance with web servers. Similarly, we can extract domain features from the headers of HTTP traffic, including the hosts with which the devices communicate.


Each domain name can be expanded from different domain levels $l$. For example, we can obtain three domain name from a three-level domain name \texttt{store.m.apple.com} from different levels, including \texttt{store.m.apple.com} ($l=3$), \texttt{m.apple.com} ($l=2$) and \texttt{apple.com} ($l=1$). For each mobile device, Magnifier counts for reconstructed domain names. Assume that $d_i^y$ is a domain name of device $y$ and $n_i^y$ is the number of $d_i^y$, the domain feature set $D^y$ is formulated as:

\begin{equation} \label{eq_bf}
	\begin{aligned}
		D^y&{= \left. \left\lbrack \left( {d_i^y,{n}_{i}^{y}} \right) \right\rbrack \right|_{i = 1}^{N^y}} \\
		&{= \left\lbrack {\left( {d_1^y,{n}_{1}^{y}} \right),\left( {d_2^y,{n}_{2}^{y}} \right),\ldots,\left( {d_{N^y}^y,{n}_{N^y}^{y}} \right)} \right\rbrack}
	\end{aligned}
\end{equation} 
where $N^y$ is the number of extracted domain names from device $y$.

\begin{figure}[t]
	\centering

	\includegraphics[width=8cm]{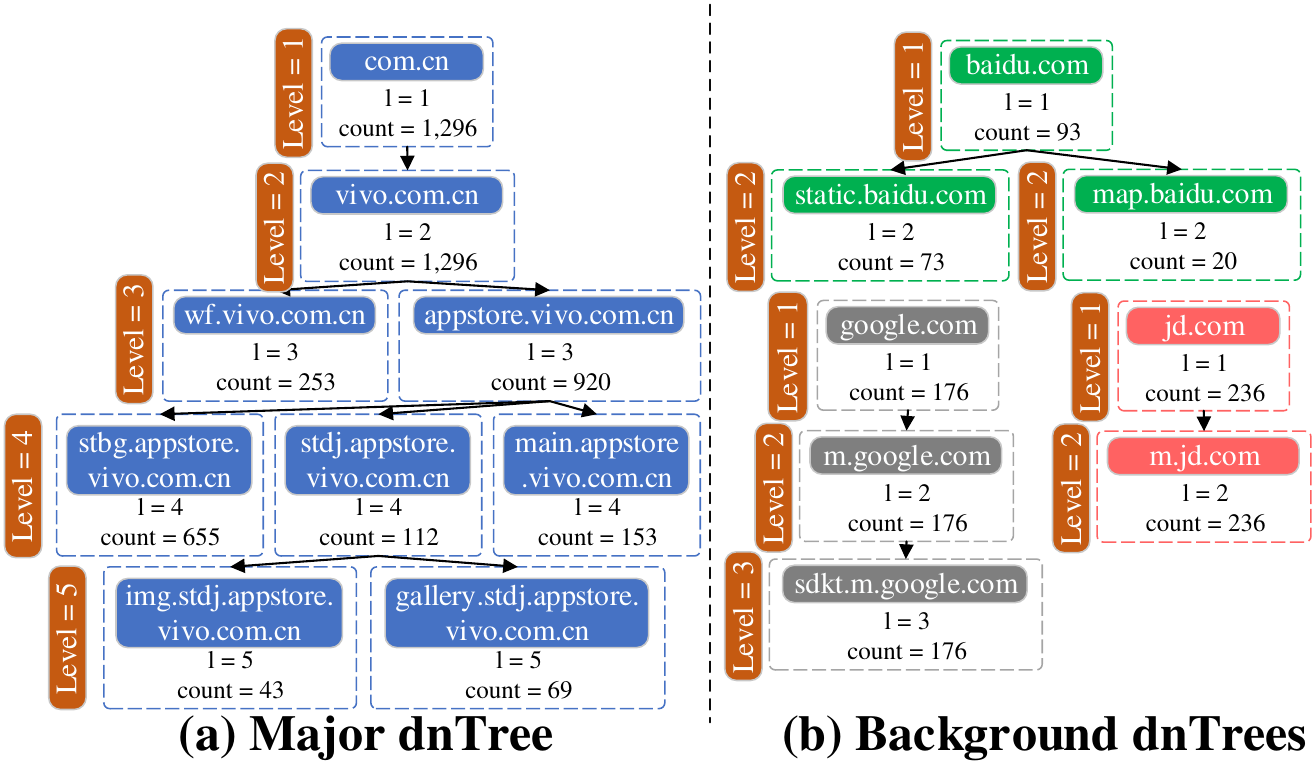}
	\caption{An example of major dnTree and background dnTrees constructed from a Vivo smart phone.}
	\label{finger_fig}

\end{figure}

\subsubsection{Initial Domain Name Forest Construction}


The conventional method of directly matching extracted domain names for network access detection often leads to high false positives. This is because online devices, such as personal computers, share domain names with the mobile devices we aim to detect. In response, Magnifier proposes a solution by modeling domain features with dnForest fingerprints, which focus on the set of accesses to domain servers.

\textbf{\emph{Definition 1.}} \textbf{dnForest. }For each mobile device class, we construct a dnForest using the domain features set $D^y$. Each dnForest comprises domain name Trees (dnTree) representing different clusters of domain features. A dnTree serves as an abstract fingerprint for a set of domain names sharing the same top-level domain (TLD). For example, a dnTree rooted by the TLD of \texttt{apple.com} contains domain names with the same TLD (i.e., \texttt{*.apple.com}), while the dnTree of \texttt{google.com} includes domain names governed by the same TLD (i.e., \texttt{*.google.com}). Importantly, there is no overlap among dnTrees within the same dnForest. dnForest of each class is formulated as:

\begin{equation}\label{dnforest}
\begin{aligned}
	{\rm dnForest^y} & =\left.\left[{\rm dnTree}^y_k\right]\right|_{k=1}^m \\
    & =\left[{\rm dnTree}^y_1,{\rm dnTree}^y_2,\ldots,{\rm dnTree}^y_m\right]
 \end{aligned}
\end{equation}
where ${\rm dnForest^y}$ is fingerprinted from device class $y$, with $m$ ${\rm dnTree}^y$.

\textbf{\emph{Definition 2.}} \textbf{dnTree. } dnTree is a Weighted Directed Acyclic Graph (WDAG), ${\rm dnTree^y_i} = (V^y_i, E^y_i, P^y_i)$, where $V^y_i$,  $E^y_i$, $P^y_i$ and $V^y_i$ refer to the node set, edge set, node property set and edge value set of ${\rm dnTree^y_i}$ in ${\rm dnTree^y}$, respectively. The node set $V^y_i$ is used to represent the domain names with the same TLD in domain feature set $D_y$, where $V^y_i = [(d_m^y,n_m^y)]|_{m=1}^{|V^y_i|} (d_{a}^y = d_{b}^y, 1 \leq a \textless b \leq |V^y_i|) $. The edge set $E^y_i$ is the correlation of two domain names (nodes) with adjacent levels. The correlation can be established between two domain names if one of them is the direct subdomain of the other. For instance, the correlation between \texttt{store.m.apple.com} ($l=3$) and \texttt{m.apple.com} ($l=2$) can be bridged while that between \texttt{store.m.apple.com} ($l=3$) and \texttt{apple.com} ($l=1$) can not. $P^y_i$ is used to record the properties of each node. Three properties are recorded in each node, including node name $d_m^y$, count $n_m^y$ of the corresponding domain name and domain level $l_m^y$. A carefully built dnTree is a tree-like graph, with a root node of TLD and sets of intermediate node as well as leaves from different domain levels. 



Magnifier formulates all domain features as a dnForest of each class of device, containing dnTrees rooted by different TLD, which is represented as the initial fingerprint of network access of each mobile device.


\subsubsection{Domain Name Forest Distillation}

The initial dnForest fingerprints all traffic generated when devices connect to the network, encompassing both specific and background traffic. Personal mobile devices host not only system-specific apps but also daily-use apps, such as those for chatting, online shopping, or entertainment. These apps connect to remote servers when the device joins the network, leading to background traffic mixing with the specific traffic. The background traffic can cause significant false alarms during classification, as these apps are commonly found on various devices. Moreover, background traffic constitutes a major portion of the captured data, leading to an imbalance among dnTrees within a dnForest. This imbalance reduces sensitivity to network access bursts.

To tackle the issues remain in the initial dnForest, Magnifier further balances the weight of dnTrees with a two-step distillation algorithm by adjusting the counts of leaves of each dnTrees. 

In a dnForest, dnTrees with more complex structures, characterized by having more decision paths, leaves, and deeper levels, theoretically indicate greater importance. This is because system-specific or brand-proprietary applications and components often communicate with a series of domains sharing the same top-level domains (TLDs), which increases the complexity of their corresponding dnTrees. In contrast, background applications are more likely to individually access domains with different TLDs once devices join a network, resulting in simpler dnTrees. As demonstrated in Figure~\ref{finger_fig}, a major dnTree (a) of the system-specific domains has more complex structure than that of background dnTrees (b).

Magnifier measures the complexity of each dnTree in a dnForest. Given a dnTree ${\rm dnTree}^y_i$ with $n$ leaves from class $y$, $n$ decision paths $\Theta^{y,i}$ can be attained:
\begin{equation}\label{decision_path}
\begin{aligned}
	\Theta^{y,i} &= \left. \left\lbrack \left( \theta_j^{y,i} \right) \right\rbrack \right|_{j = 1}^{n} \\
    \theta_j^{y,i} & = \left. \left\lbrack \left( {v_l^{y,i,j},f(v_l^{y,i,j})} \right) \right\rbrack \right|_{l = 1}^{L}
\end{aligned}
\end{equation}
where $\Theta^{y,i}$ is set of the decision paths of the $i$ dnTree of class $y$, with $n$ decision paths (leaves) $\theta_j^{y,i}$. Each decision path $\theta_j^{y,i}$ is a sequence in length (level) of $L$. $v_l^{y,i,j}$ refers to the node of the corresponding decision path in level $l$. Input with $v_l^{y,i,j}$, $f(v_l^{y,i,j})$ returns the sum of its children if $f(v_l^{y,i,j})$ is not a leaf. Otherwise, it returns the count of the node.

Then, Magnifier computes the weight of each dnTree by estimating the contribution of each decision path:
\begin{equation}\label{eq_weight}
\begin{aligned}
	W^{y,i} &=  \log{ (\sum\limits_{j=1}^{n}{ w_j^{y,i} + 1} ) } \\
    w_j^{y,i} & = \prod\limits_{l=1}^{L-1}{\log{(\frac{f(v_l^{y,i,j})}{f(v_{l+1}^{y,i,j})} + \sigma} )}
\end{aligned}
\end{equation}
where $W^{y,i}$ is the weight (complexity) of dnTree $i$ of class $y$, which is obtained by computing the contribution $w_j^{y,i}$ of each decision path. $\sigma$ is the hyper parameter for noise tolerance. Specifically, contribution $w_j^{y,i}$ of each decision path is the first-order quotient of a path, intuitively.

Magnifier updates the count of each node in each dnTree, with the corresponding weight $w_j^{y,i}$. Given the original counts $n_j^{y,i}$ of node $v_j^{y,i}$, the balanced count $\overline{n}_j^{y,i}$ can be obtained (i.e., $\overline{n}_j^{y,i} = W^{y,i} \cdot n_j^{y,i}$). The system-specific dnTrees are more likely get greater balancing weight since they have more intricate structure, which expand the intra-class scatter among dnTrees in each dnForest.

Magnifier further adjusts the inter-class scatter with a tree-based TF-IDF algorithm, to further highlight the specific fingerprint in each dnForest:
\begin{equation}
\begin{aligned}
	\hat{n}_j^{y,i} &=  {\rm TF}_j^{y,i} \cdot  {\rm IDF}_j^{y,i} \cdot \overline{n}_j^{y,i} \\
    {\rm IDF}_j^{y,i} & = \frac{\overline{n}_j^{y,i}}{ \sum\limits_{Y}\sum\limits_{V^{y^{*}}}\delta(v^{y^{*}} ==  v_j^{y,i}) } \\
    {\rm TF}_j^{y,i} & = \frac{\overline{n}_j^{y,i}}{ |V^y|} 
\end{aligned}
\end{equation}
where $Y$ is the set of the classes of devices and $y^{*} \in Y$. $V^{y^{*}}$ is the nodes in ${\rm dnForest}^{y^{*}}$. $\delta(\cdot)$ is defined as:

\begin{equation} \label{eq_jud}
\begin{aligned}
	\delta(a)\begin{cases}
			{1} & {{a = {\rm true}}} \\
			{0} & {{a = {\rm false}}} \\
		\end{cases}
\end{aligned}
\end{equation}

The proprietary dnTrees always get greater IDF since they are barely involved in the other dnForests. The two-step algorithm helps to balance the weights of dnTrees for the more rational fingerprints. 

\begin{figure}[t]
	\centering

	\includegraphics[width=8cm]{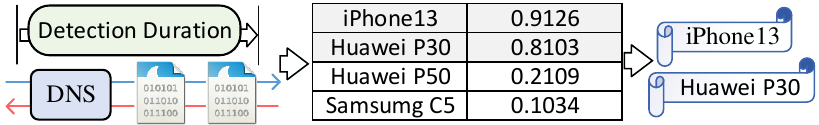}
	\caption{Real-time network access detection of Magnifier.}
	\label{fig_detect}

\end{figure}

\subsection{Real-time Network Access Detection}
As discussed, before the mobile devices establish connections with remote servers, they resolve the IP addresses of the servers through DNS requests. Based on the theory, once a DNS request is captured, Magnifier focuses on analyzing the subsequent traffic that sent by the source in $\tau$ seconds, where $\tau$ is the pre-defined detection duration, as illustrated in Figure~\ref{fig_detect}. In such way, Magnifier can be deployed on the gateway to monitor high-bandwidth network traffic efficiently by only analyzing a small part of it with limited computation resources.


Given a flow with detection duration $\tau$ (i.e., $\zeta_{\tau}$), Magnifier extracts the domain features $D'$ mentioned in Section~\ref{sec_features} (i.e., $D' = [d_i]|_{i=1}^{|D'|} = [d_1,d_2,...,d_{|D'|}]$). Then, Magnifier matches $D'$ with each dnForest of each class of mobile device to detect the behaviour of network access, which is formulated as:

\begin{equation}
\begin{aligned}
	{\rm Score^y} &= \frac{\sum\limits_{D'} \sum\limits_{F^y} \sum\limits_{\hat{V}^{y,i}} \Gamma(d_k , \hat{v}_j^{y,i}) }{ |V^y| } 
\end{aligned}
\end{equation}
where ${F^y}$ represents ${\rm dnForest}^y$ and $\hat{V}^{y,i}$ is the node set of ${\rm dnTree}^{y,i}$ in ${\rm dnForest}^y$. $\hat{v}_j^{y,i}$ is one of the nodes (domain names) of $\hat{V}^{y,i}$ (i.e., $\hat{v}_j^{y,i} \in \hat{V}^{y,i}$). $\Gamma(\cdot,\cdot)$ is defined as:
\begin{equation}
\begin{aligned}
	\Gamma(d_k , \hat{v}_j^{y,i}) \begin{cases}
			{\hat{n}_j^{y,i}} & {{d_k = \hat{v}_j^{y,i}}} \\
			{0} & {{d_k \neq \hat{v}_j^{y,i}}} \\
		\end{cases}
\end{aligned}
\end{equation}

It returns the count $\hat{n}_j^{y,i}$ of node $\hat{v}_j^{y,i}$ if the domain name of the test traffic $d_k$ matches the node. Otherwise it returns 0. Then, the test traffic $\zeta_{\tau}$ is predicted to match the fingerprint of class $y$ if the corresponding confidence ${\rm Score^y}$ is greater than the pre-defined threshold $\epsilon$ (i.e., ${\rm Score^y} \textgreater \epsilon $). Note that Magnifier tries to match $\zeta_{\tau}$ with each class of fingerprint. Therefore, Magnifier is able to detect the behaviours of network access of multiple mobile devices among one detection duration. Figure~\ref{fig_detect} shows the logic of real-time network access detection.

Due to the DNS cache in local routers, a mobile device can resolve the IP addresses without access remote DNS if it repeatedly joins the network during the Time To Live (TTL) of the DNS cache. In this scenario, less domain features can be extracted from the repetitive network access traffic compared with those from the initial traffic. In order to tackle the issues, we propose a collection mechanism, namely \emph{collector}, collecting all values of a dnTree if the coverage $\gamma$ of the dnTree is greater than the pre-defined value.


Given a set of domain features $D'$ extracted from $\zeta_{\tau}$, counts of all nodes in a dnTree $T$ can be obtained if the coverage rate is greater than the threshold $\gamma$ (i.e., $\frac{ \sum_{\hat{V}^{y,T}} \Gamma(d_k , \hat{v}_j^{y,T}) }{|V^{y,T}|} \textgreater \gamma$), where $\hat{V}^{y,T}$ is node set of major dnTree $T$ of class $y$. Note that Magnifier with \emph{collector} is able to detect both repetitive network access and initial network access. 

\section{Experiment Setup} \label{sec_expset}

\subsection{Evaluation Metrics}
Two metrics are used to evaluate the performance of our method, including Detection Rate (DR) and False Alarm Rate (FAR):

\begin{equation}
\begin{aligned}
	{\rm DR} &= \sum\limits_{(x_i,Y_i) \in S} \frac{\delta(\Psi(x_i) == Y_i)}{|S|}
\end{aligned}
\end{equation}
where $(x_i,Y_i)$ is a data-label pair in testing set $S$. $\Psi(\cdot)$ is the classification model. $\delta(\cdot)$ is defined in Eq.\ref{eq_jud}. $Y_i$ is the labels set of $x_i$ (i.e., $Y_i = [y_i^j]|_{j=1}^{K} = [y_i^1, y_i^2,..., y_i^K]$). Specially, when evaluating the performance on muliti-device network access detection (Sec. ~\ref{section_multi}), each event has more than one ground-truth labels (i.e., $|Y_i| > 1$). In rest of experiments, each event is attached to one label, where $|Y_i| = 1$.

FAR is the indicator that counts for the misclassification that predicts the networking access to background traffic: 

\begin{equation}
\begin{aligned}
	{\rm FAR} &= 1- \sum\limits_{(x^D_i, y^D) \in S^D} \frac{\delta(\Psi(x^D_i) == y^D)}{|S^D|}
\end{aligned}
\end{equation}
where $(x^D_i, y^D)$ is a testing sample in background event set $S^D$. $y^D$ is the label of background events.

\subsection{Comparison Baselines}
To the best of our knowledge, there have been limited methods proposed for the detection of mobile device network access behavior. However, related work that focuses on encrypted network classification shares some common concepts. As a result, we compare Magnifier with these methods after fine-tuning.

Inspired by~\cite{prograph}, we re-implement four types of state-of-the-art Encrypted Traffic Classification (ETC) methods, including statistical models (RF~\cite{randomforest}, C4.5~\cite{decisiontree}, and XGBoost~\cite{xgboost}), deep learning models (FS-Net~\cite{fsnet} and MBTree~\cite{mbtree}), fingerprint models (FlowPrint~\cite{flowprint} and ProGraph~\cite{prograph}), and end-to-end models (CNN-RNN~\cite{cnnrnn} and FC-Net~\cite{fcnet}). It's worth noting that these baselines analyze network traffic for each detection duration during prediction, which aligns with the methodology employed by Magnifier.

\begin{figure}[t]
	\centering

	\includegraphics[width=8cm]{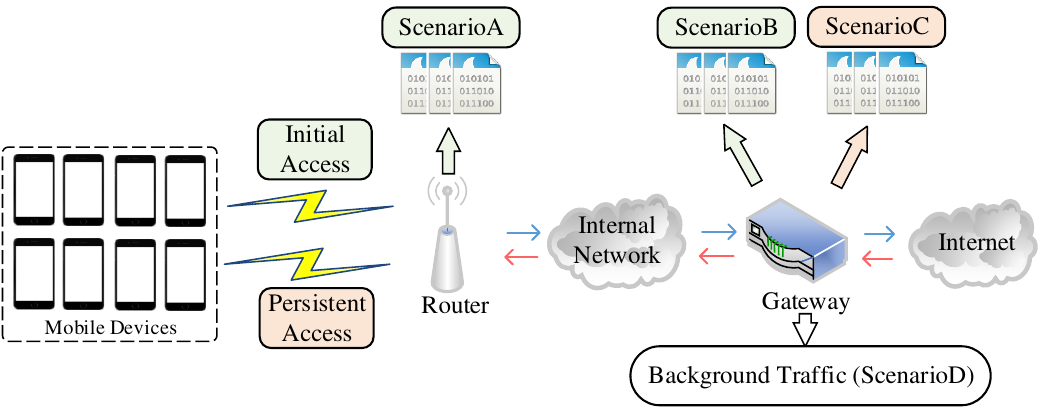}
	\caption{Traffic capturing in real-world scenarios of NetCess2023.}
	\label{fig_dataset}

\end{figure}

\subsection{Dataset Organization}

To assess the effectiveness of Magnifier, we meticulously curated a dataset known as NetCess2023. This dataset comprises network access traffic from 26 distinct types of mobile devices representing 7 different brands. We encompass 4 real-world scenarios, encompassing three network access scenarios and one background traffic scenario.

As depicted in Figure~\ref{fig_dataset}, the traffic is captured at various network locations. In ScenarioA, we deploy a sniffer on a local router, allowing us to capture the pure traffic of network access as soon as a device joins the network. Notably, the traffic in ScenarioA exclusively involves the access behavior of the target devices. Consequently, ScenarioA serves both for Magnifier's fingerprinting and as the training set for the baseline models.

In ScenarioB, ScenarioC, and ScenarioD, we capture traffic at one of the backbone gateways within our campus network. ScenarioB encompasses the traffic associated with the initial network access, which includes the complete network access traffic. In contrast, ScenarioC captures repetitive network traffic, including only partial traffic from each network access, primarily due to the DNS cache in the local routers. It's important to note that background traffic is mixed in both ScenarioB and ScenarioC.

In ScenarioD, we capture traffic that does not contain any network access behaviors from the included mobile devices.

NetCess2023 is a carefully designed dataset with about 10GB raw network traffic from 4 real-world scenarios. It collects the traffic data of network accessing of 26 different mobile devices models from 6 brands, covering most of the mainstream mobile phones. This dataset comprises approximately 10GB of network traffic, encompassing a broad spectrum of mainstream mobile device brands, including Apple (5 models), Huawei (5 models), Samsung (2 models), Xiaomi (5 models), Vivo (5 models), Oppo (3 models), and Coolpad (1 model). Each time a device joins the network, the traffic can be automatically captured and recorded as an event (Evt). Four Scenarios are investigated. About 2,585 events are included in ScenarioA. More than 1,400 events and 2,400 events are recorded in ScenarioB and ScenarioC, respectively. ScenarioD, which is not shown in the table, contains only background traffic without the behaviour of network accessing. NetCess2023 has been made publicly available\footnote{https://github.com/SecTeamPolaris/Magnifier/tree/ main/NetCess2023}

\subsection{Evaluation of Baselines} \label{baseline_sec}
Since the lack of methods that focus on network accessing detection, we have to reconstruct and fine-tune the analogous models. Specifically, we train or initialize the baselines with ScenerioA of NetCess2023. Each event in ScenarioA is considered as a single training sample. DR and FAR are obtained with two-stage validation. Firstly, testing events in ScenarioB or ScenarioC are predicted, by which DR can be calculated. Then, ScenarioD, which contains only background traffic, is used to validate the FAR of each baseline.

Learning-based methods such as RF, FS-Net, FC-Net or CNN-RNN follow the same way to perform initialization and detection as mentioned in their proposal. However, we re-design the graph initialization of ProGraph since clustering with IP features can hardly result in a legitimate graph. Specifically, each event is considered as a node during initialization so that features and labels can be aggregated and disseminated as expected.

\begin{table}[t]  
	\caption{The organization of NetCess2023. NetCess2023 investigates the traffic of network access from 26 mobile devices, covering 7 brands of the mainstream brands.} 
	\label{tab_dataset}
	\centering   
	\setlength\tabcolsep{3pt}

	\begin{tabular}{clcccccc}  
		\toprule
		&&\multicolumn{2}{c}{ScenarioA}&\multicolumn{2}{c}{ScenarioB}&\multicolumn{2}{c}{ScenarioC}\cr

        \cmidrule(r){3-4} \cmidrule(r){5-6} \cmidrule(r){7-8}

        Brands&Models&Bytes&Evts&Bytes&Evts&Bytes&Evts\cr
		\midrule
		\multirow{5}{*}{\rotatebox{0}{Apple}}&iPhone13U& 6.25M & 72 & 3.19M & 35 & 12.2M & 73\cr
        &iPhoneXR& 5.82M & 90 & 6.12M & 60 & 16.4M & 126 \cr
        &iPhone7& 7.70M & 91 & 3.79M & 41 & 6.97M & 92 \cr
        &iPhone6s& 38.4M & 90 & 22.1M & 26 & 38.1M & 32 \cr
        &iPhone6& 126M & 90 & 51.9M & 59 & 684M & 75 \cr

        \midrule
		\multirow{5}{*}{\rotatebox{0}{Huawei}}&Mate50& 29.4M & 90 & 17.5M & 54 & 550M & 60\cr
        &Mate40P& 18.2M & 90 & 43.6M & 59 & 849M & 79 \cr
        &Mate30& 10.1M & 89 & 3.63M & 37 & 24.0M & 107 \cr
        &P30& 37.1M & 90 & 16.1M & 56 & 10.2M & 15 \cr
        &P9& 26.6M & 97 & 11.9M & 65 & 48.9M & 115 \cr

        \midrule
		\multirow{5}{*}{\rotatebox{0}{Xiaomi}}&Xiaomi13U& 17.6M & 79 & 11.4M & 53 & 15.5M & 29\cr
        &Xiaomi6x& 12.3M & 90 & 8.31M & 60 & 23.0M & 150 \cr
        &XiaomiMix2& 32.5M & 88 & 9.60M & 46 & 15.5M & 28 \cr
        &Xiaomi3& 10.7M & 151 & 1.87M & 73 & 744M & 141 \cr
        &Xiaomi2s& 2.28M & 90 & 1.44M & 56 & 1.21G & 132 \cr

        \midrule
		\multirow{5}{*}{\rotatebox{0}{Vivo}}&VivoX30& 9.74M & 90 & 4.70M & 42 & 8.13M & 108\cr
        &VivoX27& 36.2M & 73 & 7.79M & 41 & 20.8M & 87 \cr
        &VivoX23& 32.7M & 133 & 22.6M & 86 & 61.1M & 213 \cr
        &VivoY66i& 11.1M & 85 & 7.14M & 44 & 35.6M & 134 \cr
        &Vivo93& 6.18M & 90 & 54.8M & 60 & 30.1M & 45 \cr

        \midrule
		\multirow{3}{*}{\rotatebox{0}{Oppo}}&OppoR15& 10.2M & 135 & 4.46M & 75 & 2.07G& 166\cr
        &OppoR9m& 9.60M & 90 & 8.97M & 58 & 29.1M & 136 \cr
        &OppoA5& 7.06M & 89 & 2.08M & 59 & 29.2M & 135 \cr

        \midrule
		\multirow{2}{*}{\rotatebox{0}{Samsung}}&SamsungS10& 5.04M & 113 & 2.95M & 47 & 57.8M & 121\cr
        &SamsungC5& 10.7M & 214 & 2.29M & 104 & 23.1M & 24 \cr

        \midrule
		\multirow{1}{*}{\rotatebox{0}{Coolpad}}&Dazen& 26.2M & 86 & 3.34M & 45 & 203M & 69 \cr

		\bottomrule
		
	\end{tabular} 
	
\end{table}

\section{Experimental Results} \label{sec_expres}
The experiments are evaluated using the following platforms: Intel i7-9750 @2.6GHz, 16GB RAM, NVIDIA GeForce RTX2060 and Windows 11. Learning-based baselines are evaluated under Python 3.11.4, Sklearn 1.3.0, CUDA 10.1, and PyTorch 1.0.1 with the default parameters. In the following experiments, the parameters are evaluated as $\sigma = 1$, $\epsilon = 0.4$ and $\gamma = 0.5$. 

In this section, we aim to answer the following questions:

\begin{itemize}
	\item How does Magnifier perform on detecting the initial behaviour of network access? (Sec. \ref{section_inital})
	\item How does Magnifier perform on detecting the repetitive behaviour of network access? (Sec. \ref{section_rep})
	\item How does Magnifier perform on multi-device network access detection? (Sec. \ref{section_multi})
	\item How efficient is Magnifier when deploying for real-time detection practically? (Sec. \ref{section_com})
	\item How does each part of the algorithm contribute to Magnifier? (Sec. \ref{section_abl})
\end{itemize}

\begin{table*}[!t]

\renewcommand{\arraystretch}{1.0}
	\caption{Experimental results on detecting initial (Section~\ref{section_inital}) and repetitive (Section~\ref{section_rep}) network access of mobile devices.} 
	\label{fig_res1}
	\centering   
	
	\setlength\tabcolsep{3.1pt}
	
	\begin{tabular}{clcccccccccccc}  
		\toprule
		&&\multicolumn{6}{c}{Brand-Level Classification (7+1 classes)}&\multicolumn{6}{c}{Model-Level Classification (26+1 classes)}\cr
		\cmidrule(r){3-8} \cmidrule(r){9-14} 
        &&\multicolumn{2}{c}{$\tau = 5$ ($s$)}&\multicolumn{2}{c}{$\tau = 10$ ($s$)}&\multicolumn{2}{c}{$\tau = 15$ ($s$)}&\multicolumn{2}{c}{$\tau = 5$ ($s$)}&\multicolumn{2}{c}{$\tau = 10$ ($s$)}&\multicolumn{2}{c}{$\tau = 15$ ($s$)}\cr
        \cmidrule(r){3-4} \cmidrule(r){5-6} \cmidrule(r){7-8}  \cmidrule(r){9-10} \cmidrule(r){11-12} \cmidrule(r){13-14} 
		&&{\bfseries DR (\%)}&{\bfseries FAR (\%)}&{\bfseries DR (\%)}&{\bfseries FAR (\%)}&{\bfseries DR (\%)}&{\bfseries FAR (\%)}&{\bfseries DR (\%)}&{\bfseries FAR (\%)}&{\bfseries DR (\%)}&{\bfseries FAR (\%)}&{\bfseries DR (\%)}&{\bfseries FAR (\%)}\cr
  
		\midrule
		\multirow{10}{*}{\rotatebox{90}{$S_A$ / $S_B$ + $S_D$ (Sec. \ref{section_inital})}}&RF \cite{randomforest}& 49.51 & 14.73 & 59.60 & 13.09 & 62.78 & 12.97 & 27.41 & 37.70 & 47.10 & 25.96 & 55.43 & 25.50 \cr
        &C4.5 \cite{decisiontree}& 32.09 & 23.36 & 35.28 & 19.95 & 35.43 & 18.57 & 20.76 & 59.65 & 26.82 & 53.36 & 25.46 & 49.93 \cr
        &XGBoost \cite{xgboost}& 51.07 & 11.77 & 62.79 & 10.43 & 67.52 & 9.24 & 42.76 & 29.07 & 53.10 & 24.79 & 59.93 & 22.87 \cr
        &FS-Net \cite{fsnet}& 78.20 & 10.92 & 87.29 & 8.07 & 93.77 & 7.48 & 65.00 & 14.63 & 83.93 & 13.87 & 91.06 & 13.62 \cr
        &MBTree \cite{mbtree}& 76.71 & 10.88 & 78.02 & 9.45 & 84.73 & 9.27 & 66.28 & 19.82 & 75.79 & 15.86 & 77.04 & 15.71 \cr
        &FlowPrint \cite{flowprint}& \underline{82.79} & 6.73 & \underline{92.47} & \underline{5.24} & \underline{96.14} & \underline{2.98} & \underline{70.81} & 9.12 & \underline{86.71} & 5.11 & \underline{95.02} & \underline{3.82} \cr
        &ProGraph \cite{prograph}& 79.84 & \underline{5.23} & 89.07 & 6.11 & 91.74 & 6.86 & 69.01 & \underline{3.81} & 83.65 & \underline{4.50} & 90.19 & 4.62 \cr
        &CNN-RNN \cite{cnnrnn}& 54.73 & 31.84 & 63.92 & 26.71 & 65.87 & 24.00 & 39.91 & 36.72 & 45.14 & 35.87 & 47.76 & 33.31 \cr
        &FC-Net \cite{fcnet}& 78.10 & 7.21 & 84.51 & 6.82 & 85.88 & 6.60 & 64.73 & 8.27 & 77.31 & 7.24 & 85.60 & 6.98 \cr
        &\textbf{Magnifier} & \textbf{85.56} & \textbf{0.08} & \textbf{93.16} & \textbf{0.15} & \textbf{98.41} & \textbf{0.25} & \textbf{73.80} & \textbf{0.08} & \textbf{89.21} & \textbf{0.15} & \textbf{97.16} & \textbf{0.25} \cr

        \midrule
        \multirow{10}{*}{\rotatebox{90}{$S_A$ / $S_C$ + $S_D$ (Sec. \ref{section_rep})}}&RF \cite{randomforest}& 45.87 & 15.41 & 53.81 & 13.77 & 55.60 & 11.93 & 24.90 & 39.09 & 29.79 & 26.11 & 34.81 & 23.89 \cr
        &C4.5 \cite{decisiontree}& 31.37 & 21.81 & 33.96 & 19.94 & 34.01 & 18.83 & 12.89 & 57.16 & 21.75 & 55.47 & 22.51 & 53.05 \cr
        &XGBoost \cite{xgboost}& 47.37 & 15.17 & 59.87 & 14.79 & 62.20 & 14.84 & 41.59 & 35.80 & 49.33 & 27.36 & 51.91 & 26.39 \cr
        &FS-Net \cite{fsnet}& 71.45 & 14.52 & 79.92 & 14.10 & 82.51 & 12.89 & 62.07 & 19.15 & 67.89 & 12.87 & 72.96 & 11.90 \cr
        &MBTree \cite{mbtree}& 74.98 & 13.78 & 75.90 & 12.42 & 79.00 & 11.91 & \underline{65.76} & 21.54 & 74.08 & 16.47 & 75.16 & 14.79 \cr
        &FlowPrint \cite{flowprint}& 76.29 & 13.87 & \underline{88.40} & 7.62 & \underline{95.76} & \underline{4.03} & 61.70 & 14.29 & 79.59 & 8.94 & \underline{93.12} & 6.95 \cr
        &ProGraph \cite{prograph}& 77.86 & \underline{4.12} & 85.39 & \underline{4.25} & 88.07 & 5.97 & 63.95 & \underline{2.90} & 76.73 & \underline{3.56} & 85.20 & \underline{3.69} \cr
        &CNN-RNN \cite{cnnrnn}& 53.91 & 32.12 & 59.34 & 29.41 & 60.87 & 28.96 & 37.84 & 40.24 & 39.63 & 36.16 & 41.67 & 35.59 \cr
        &FC-Net \cite{fcnet}& \underline{79.83} & 9.29 & 82.49 & 9.10 & 83.52 & 7.37 & 60.59 & 11.56 & \underline{79.61} & 10.14 & 81.04 & 9.57 \cr
        &\textbf{Magnifier} & \textbf{80.46} & \textbf{0.08} & \textbf{90.31} & \textbf{0.15} & \textbf{97.54} & \textbf{0.25} & \textbf{69.07} & \textbf{0.08} & \textbf{83.53} & \textbf{0.15} & \textbf{96.83} & \textbf{0.25} \cr

		\bottomrule
		
	\end{tabular} 

\end{table*}

\subsection{Detecting Initial Network Access of Mobile Devices} \label{section_inital}

We assess the detection of initial network access for the mobile devices included in NetCess2023. Our evaluation, as shown in Table~\ref{fig_res1}, is conducted in two scenarios. Brand-level classification aims to differentiate between the brands of mobile devices (e.g., Apple, Huawei, or Samsung), resulting in an 8-class classification task (7 brands of mobile devices and 1 class for background network). Model-level classification, on the other hand, detects network access at the model level (e.g., iPhone13U, iPhoneXR, SamsungS10, or HuaweiP30), which corresponds to a 27-class classification (26 device models and 1 class for background network).

When detecting initial network access behavior, models are initialized/trained using traffic from ScenarioA ($S_A$) and validated using a combination of ScenarioB ($S_B$) and ScenarioD ($S_D$). Additionally, models are evaluated under various settings of the detection duration ($\tau$), with detection durations set to 5$s$, 10$s$ and 15$s$ respectively.


As shown in Table~\ref{fig_res1}, Magnifier achieves eye-catching performance when detecting the initial network access in both of brand-level and model-level classification with 85.56\% and 73.80\% of DR and 0.08\% of FAR, which is 2.77\% higher and 6.65\% lower than those of FlowPrint~\cite{flowprint} when $\tau = 5$. Magnifier obtains the best performance when $\tau = 15s$ and remains stable despite the increase of the detection duration. This is because most of the behaviour of network access lasts no more than 15 seconds. 

\subsection{Detecting Repetitive Network Access of Mobile Devices}\label{section_rep}
Detecting repetitive network access is likely to be more challenging since the local DNS cache may block part of the access traffic. As shown in Table~\ref{fig_res1}, models are initialized/trained with $S_A$ and validated with the combination of $S_C$ and $S_D$. Similarly, they undergo evaluation in two classification levels, under different setting of detection duration $\tau$. Due to the deficiency of domain features, existing models suffer from significant performance degradation. The DR of FS-Net~\cite{fsnet} drops from 91.06\% to 72.96\% while ProGraph~\cite{prograph} drops from 90.19\% to 85.20\%, under the setting of $\tau = 15s$ on 27-class classification. As discussed, collector mechanism is proposed to counter the feature deficiency. Therefore, the DR of Magnifier slightly drops by 0.33\% in $S_C$, compared with that of in $S_B$.

\begin{table}[!t]

\renewcommand{\arraystretch}{0.9}
	\caption{Network access detection in multi-device scenario. DR with multi-labels is presented.} 
	\label{fig_res2}
	\centering   
	
	\begin{tabular}{clcccc}  
		\toprule
		&&\multicolumn{2}{c}{Brand Level}&\multicolumn{2}{c}{Model Level}\cr
		\cmidrule(r){3-4} \cmidrule(r){5-6} 
        &&$S_B (\%)$&$S_C (\%)$&$S_B (\%)$&$S_C (\%)$\cr
		\midrule
  \multirow{3}{*}{\rotatebox{90}{$K=2$ }}&\textbf{Magnifier} ($\tau=5$) & 84.22 & 72.55 & 67.84 & 66.14 \cr
        &\textbf{Magnifier} ($\tau=10$) & 90.82 & 83.51 & 83.19 & 81.22 \cr
        &\textbf{Magnifier} ($\tau=15$) & 96.26 & 94.20 & 96.54 & 91.71 \cr
        
        \midrule

        \multirow{3}{*}{\rotatebox{90}{$K=3$ }}&\textbf{Magnifier} ($\tau=5$) & 82.15 & 74.40 & 56.02 & 46.16 \cr
        &\textbf{Magnifier} ($\tau=10$) & 88.14 & 83.86 & 77.03 & 75.23 \cr
        &\textbf{Magnifier} ($\tau=15$) & 96.41 & 92.86 & 94.60 & 84.41 \cr

		\bottomrule
		
	\end{tabular} 

\end{table}

\subsection{Multi-device Network Access Detection}\label{section_multi}
During the same detection duration, there may be more than one devices join the network. However, few methods on ETC claim the ability to detect the multiple device network access since they generally make single prediction on each traffic flow. Magnifier, can tackle the issue because they subsequently match the traffic with fingerprint of each class and make the prediction independently. Note that detection threshold ($\epsilon$) and coverage ($\gamma$) are set to 0.4 and 0.5, respectively.

We reconstruct traffic in ScenarioB and ScenarioC respectively by combining the network access traffic of two ($K=2$) or three ($K=3$) different classes. Specifically, we combine traffic of different brands on brand-level classification while combining traffic of different models on model-level classification. The experimental results on DR are shown in Table~\ref{fig_res2}. As discussed, a prediction is considered to be correct only when all classes of the test traffic are precisely predicted. Magnifier achieves 96.26\% and 94.20\% of DR on brand-level classification in ScenarioB ($S_B$) and ScenarioC ($S_C$) respectively, where $K=2$ and $\tau = 15s$. When $k$ increases to 3, the DR drops to 96.41\% and 92.86\%. Comparatively, the DR of  Magnifier slightly drops on model-level classification. This is because the multi-device traffic mixes similar fingerprint from the same brand.

\subsection{Evaluation of Computational Consumption}\label{section_com}
\begin{figure}[t]
	\centering

	\subfloat{\includegraphics[width=8.5cm]{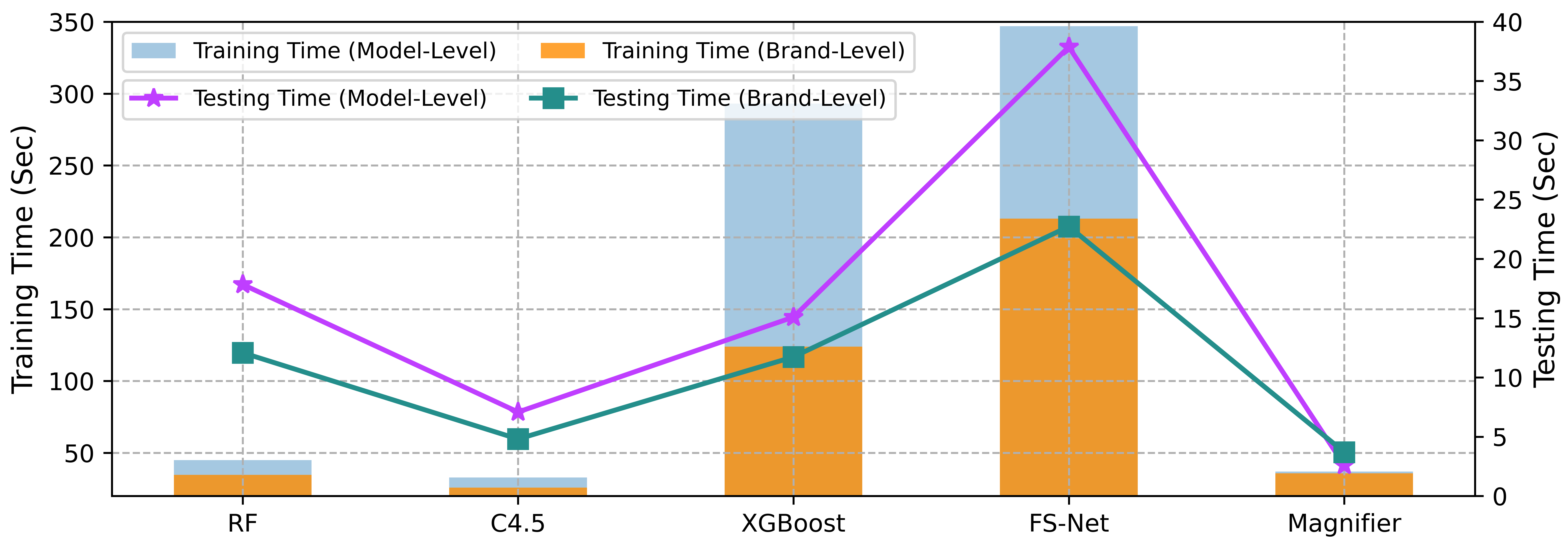}} \\
	\subfloat{\includegraphics[width=8.5cm]{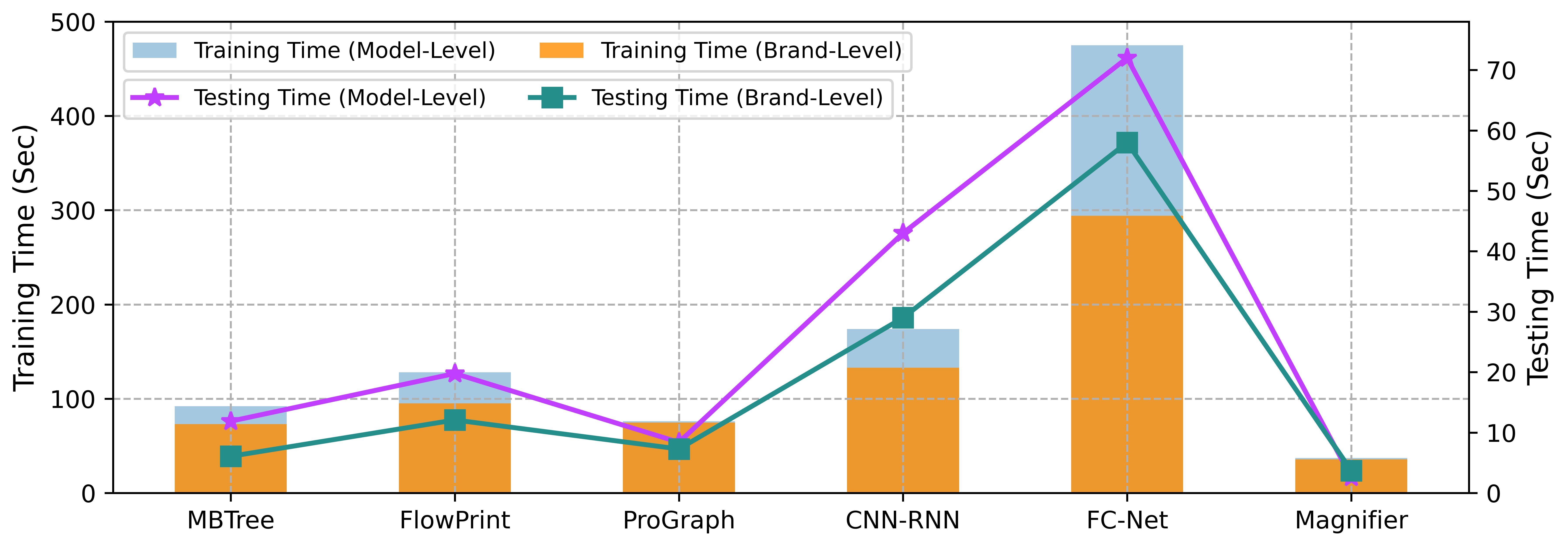}}
	\caption{Comparison on computational consumption.}
	\label{fig_consum}

\end{figure}

We evaluate the time consumption on model training and validation. Models are trained/initialized with $S_A$ and validated with $S_B + S_D$, under the setting of $\tau = 15s$. Figure~\ref{fig_consum} shows the results on both brand-level and model-level classification. As demonstrated, Magnifier spends similar time on brand-level (35s) and model-level (37s) fingerprinting. This is because Magnifier has to fingerprint each model preliminarily on both brand level and model level. Magnifier spends 3.7s and 2.6s when testing with $S_B + S_D$ on brand level and model level respectively, 
and testing when classifying different brands, which outperforms most learning-based methods. Empowered by the proposed fingerprinting algorithms, Magnifier can promise both effectiveness and efficiency of real-time network access detection on lightweight deployment practically.

\begin{table}[t]

    \renewcommand{\arraystretch}{1.5}
	\caption{Ablation study on Magnifier.} 
	\label{tbl_abl1}
	\centering   

	\setlength\tabcolsep{3.1pt}
    
	\begin{tabular}{ccccccc}  
		\toprule
		&&&\multicolumn{2}{c}{Brand Level}&\multicolumn{2}{c}{Model Level}\cr
		\cmidrule(r){4-5} \cmidrule(r){6-7} 
        &Distillation&Collector&$\textbf{DR} (\%)$&$\textbf{FAR}  (\%)$&$\textbf{DR}  (\%)$&$\textbf{FAR}  (\%)$\cr
		\midrule
        \multirow{4}{*}{\rotatebox{90}{$S_A$ / $S_B$ + $S_D$}}&\textcolor{red!100}{\XSolidBrush}& \textcolor{red!100}{\XSolidBrush} & 52.77 & 9.02 & 48.33 & 9.02 \cr
        &\textcolor{red!100}{\XSolidBrush}&\textcolor{green!100}{\CheckmarkBold} & 72.50 & 13.73 & 67.97 & 13.73 \cr
        &\textcolor{green!100}{\CheckmarkBold}&\textcolor{red!100}{\XSolidBrush} & 96.83 & \textbf{0.03} & 94.19 & \textbf{0.03} \cr
        &\textcolor{green!100}{\CheckmarkBold}& \textcolor{green!100}{\CheckmarkBold}  & \textbf{98.41} & 0.25 & \textbf{97.16} & 0.25 \cr

        \midrule
        \multirow{4}{*}{\rotatebox{90}{$S_A$ / $S_C$ + $S_D$}}&\textcolor{red!100}{\XSolidBrush}& \textcolor{red!100}{\XSolidBrush} & 40.83 & 9.02 & 31.10 & 9.02 \cr
        &\textcolor{red!100}{\XSolidBrush}&\textcolor{green!100}{\CheckmarkBold} & 47.79 & 13.73 & 43.98 & 13.73 \cr
        &\textcolor{green!100}{\CheckmarkBold}&\textcolor{red!100}{\XSolidBrush} & 83.92 & \textbf{0.03} & 78.57 & \textbf{0.03} \cr
        &\textcolor{green!100}{\CheckmarkBold}& \textcolor{green!100}{\CheckmarkBold}  & \textbf{97.54} & 0.25 & \textbf{96.83} & 0.25 \cr
	
		\bottomrule
		
	\end{tabular} 

\end{table}

\subsection{Ablation Study}\label{section_abl}
We evaluate the contribution of each components of Magnifier. As discussed, two enhancement empowers the initial Magnifier, including the distillation algorithm and the mechanism of collector. Here, Magnifier is equipped with distillation and collector respectively to verify the contribution of each component. As illustrated in Table~\ref{tbl_abl1}, Magnifier is evaluated in two scenarios (i.e., $S_A$ / $S_B$ + $S_D$ and $S_A$ / $S_C$ + $S_D$). Specifically, detection duration $\tau$ is set to 15s while $\epsilon = 0.4$ and $\gamma = 0.5$. Magnifier without distillation achieves 67.97\% of DR and 13.73\% of FAR. The fingerprints without distillation mistakenly focus on dnTrees of background traffic, which leads to unacceptable detection accuracy and false alarms. Although the FAR of Magnifier achieves poorer FAR (0.25\%) than the one without the mechanism of collector (0.03\%), it promises superior DR (97.16\%) than Magnifier without collector (94.16\%). This is because the values of the background dnTrees may also be collected when auditing background traffic, which results in false alarms. As discussed, collector mechanism is proposed to counter the fragmentary features brought by local DNS cache. Therefore, Magnifier with collector performs much better when evaluating in $S_C$, of which the DR improves by 18.26\% (from 78.57\% to 96.83\%). 

The detection threshold $\epsilon$ and coverage rate $\gamma$ are further discussed. The value of $\epsilon$ brings different trade-off between DR and FAR. Higher $\epsilon$ brings better FAR but poorer DR and vice versa. It shows that Magnifier achieves the best performance when $\epsilon = 0.4$ and $\gamma = 0.5$. Too higher of $\gamma$ will suffer from false negative while the opposite may bring out too much false alarms.

\begin{figure*}[t]
	\centering
	
	\subfloat[$\gamma = 0.1$]{\includegraphics[width=4.5cm]{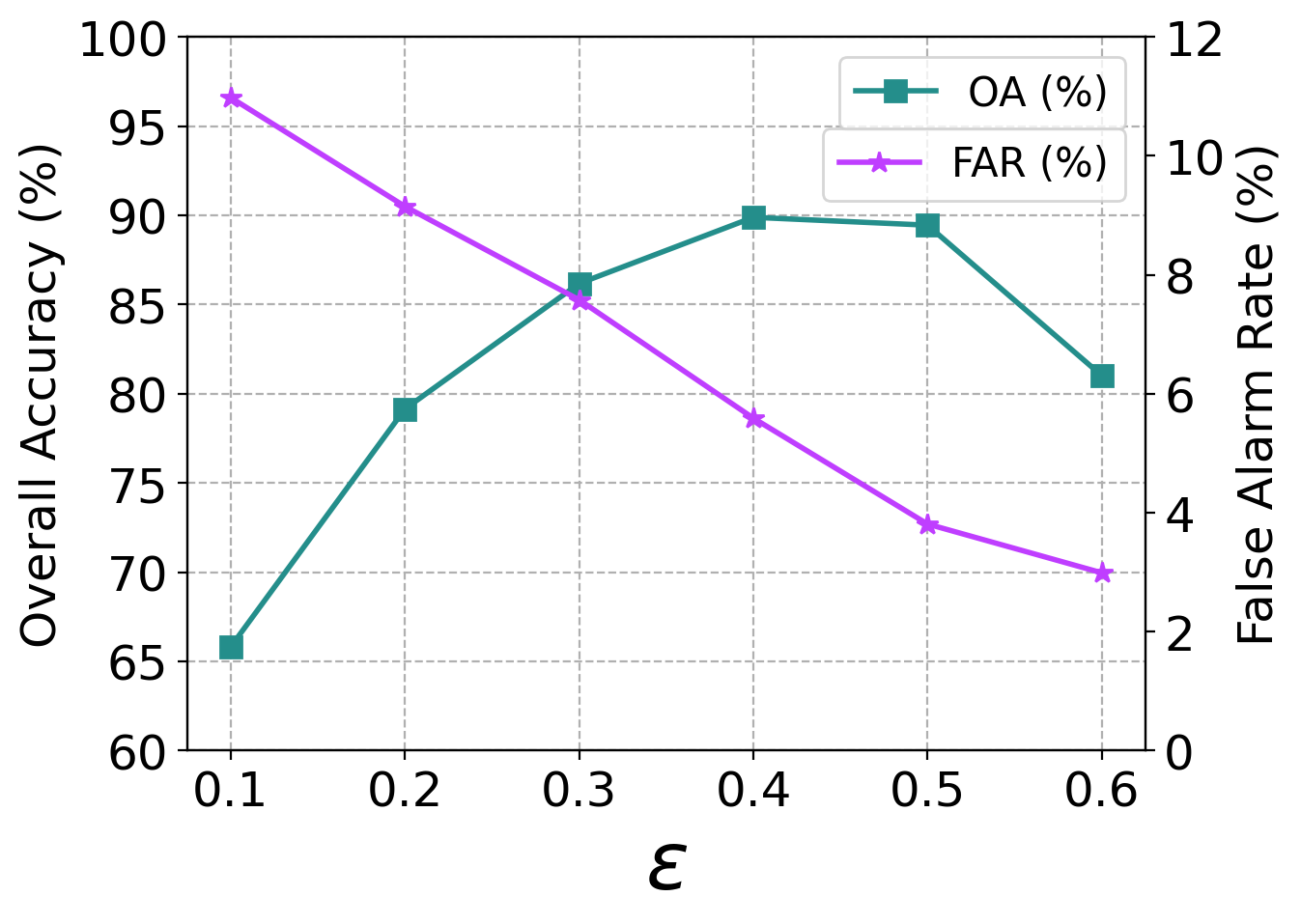}} \subfloat[$\gamma = 0.2$]{\includegraphics[width=4.5cm]{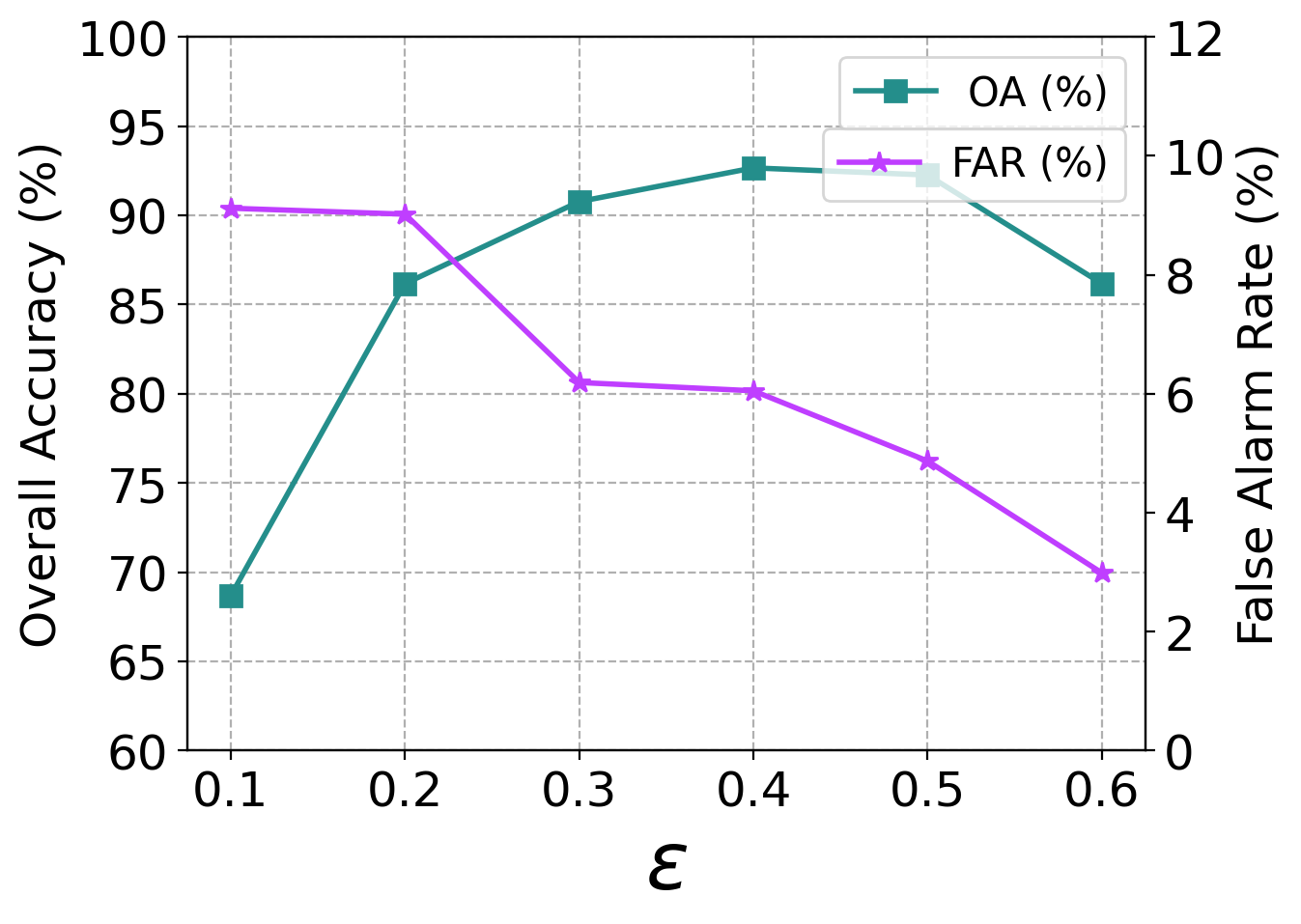}} \subfloat[$\gamma = 0.3$]{\includegraphics[width=4.5cm]{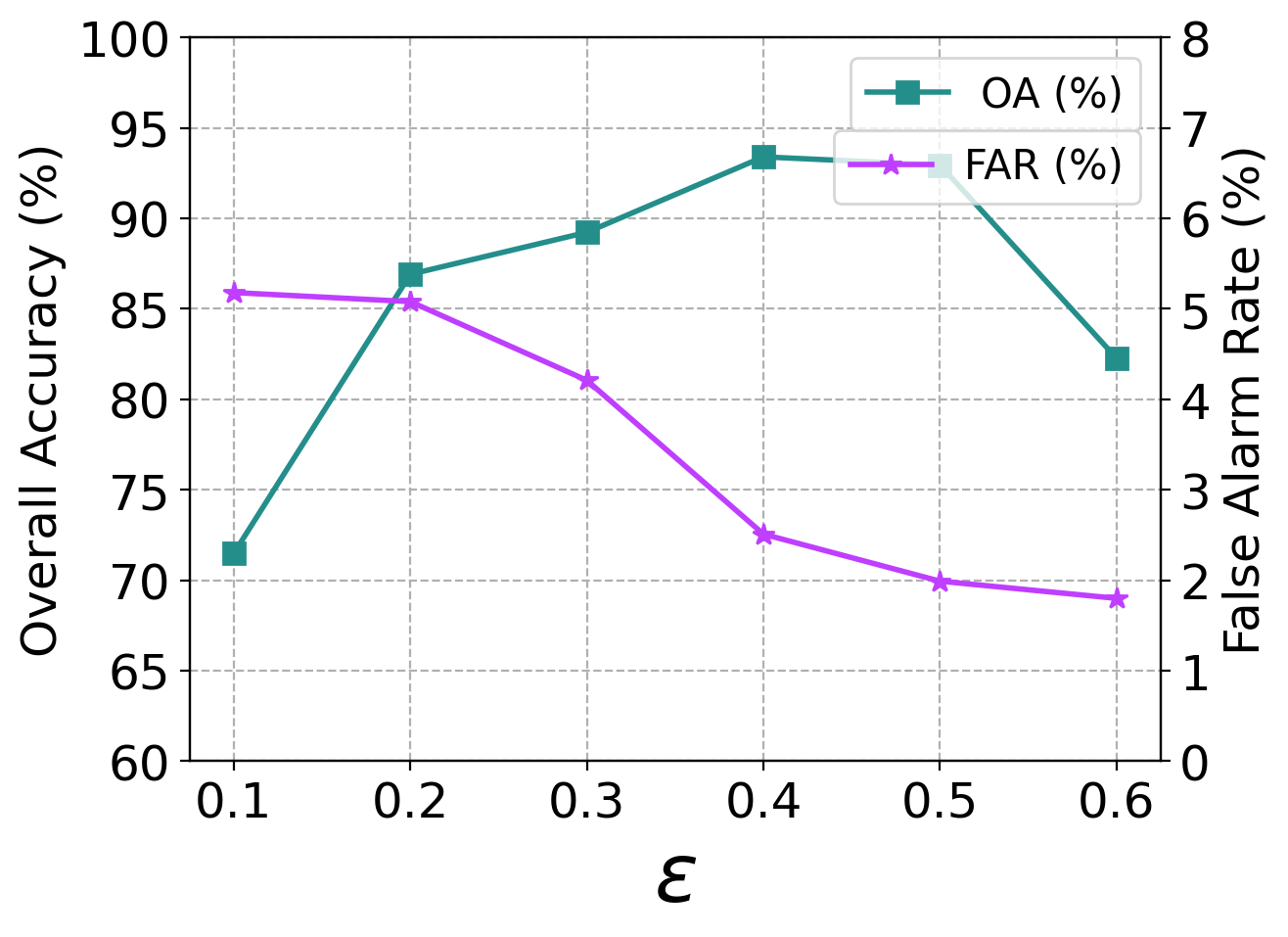}} \subfloat[$\gamma = 0.4$]{\includegraphics[width=4.5cm]{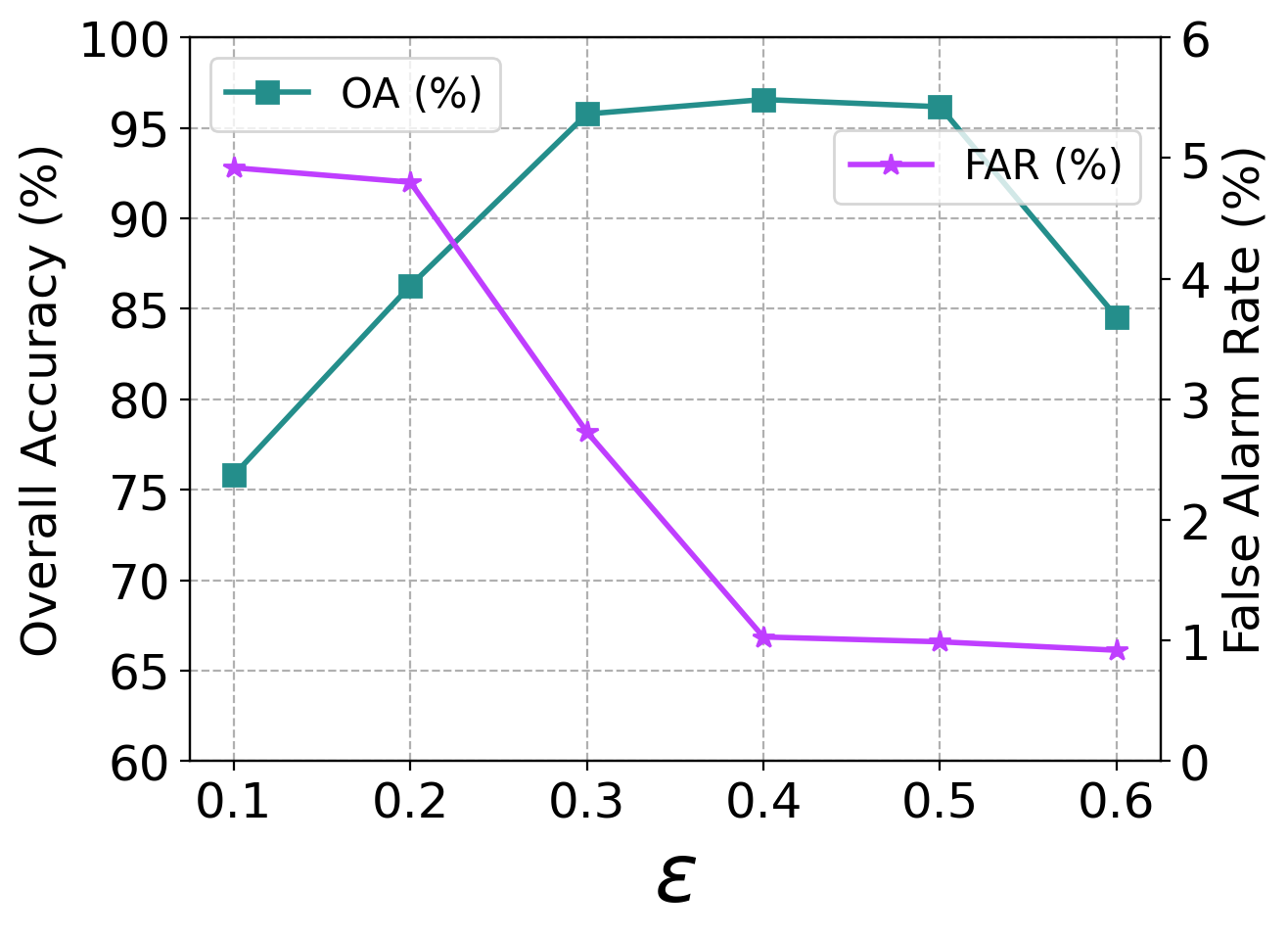}}\\
    \subfloat[$\gamma = 0.5$]{\includegraphics[width=4.5cm]{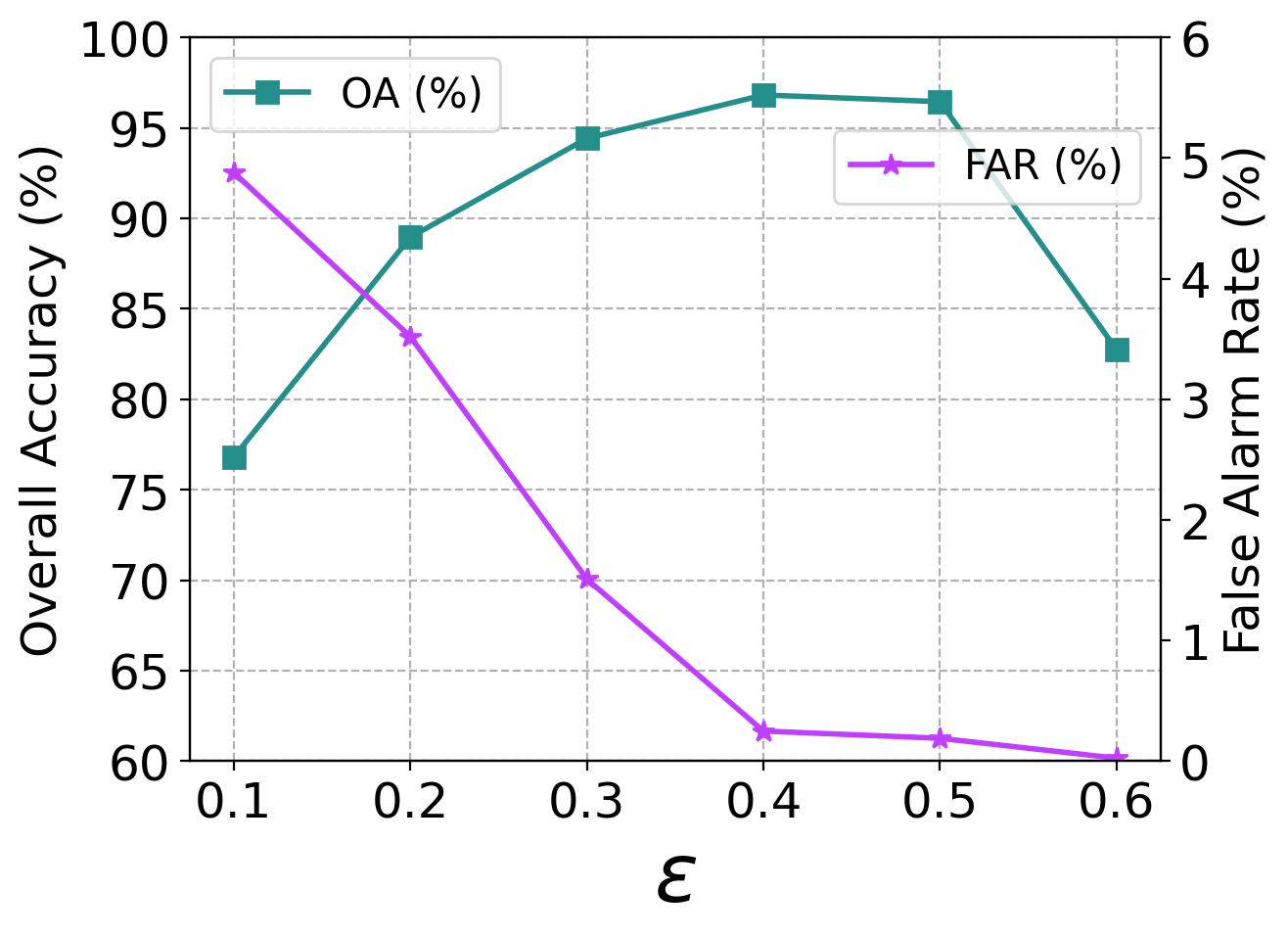}} \subfloat[$\gamma = 0.6$]{\includegraphics[width=4.5cm]{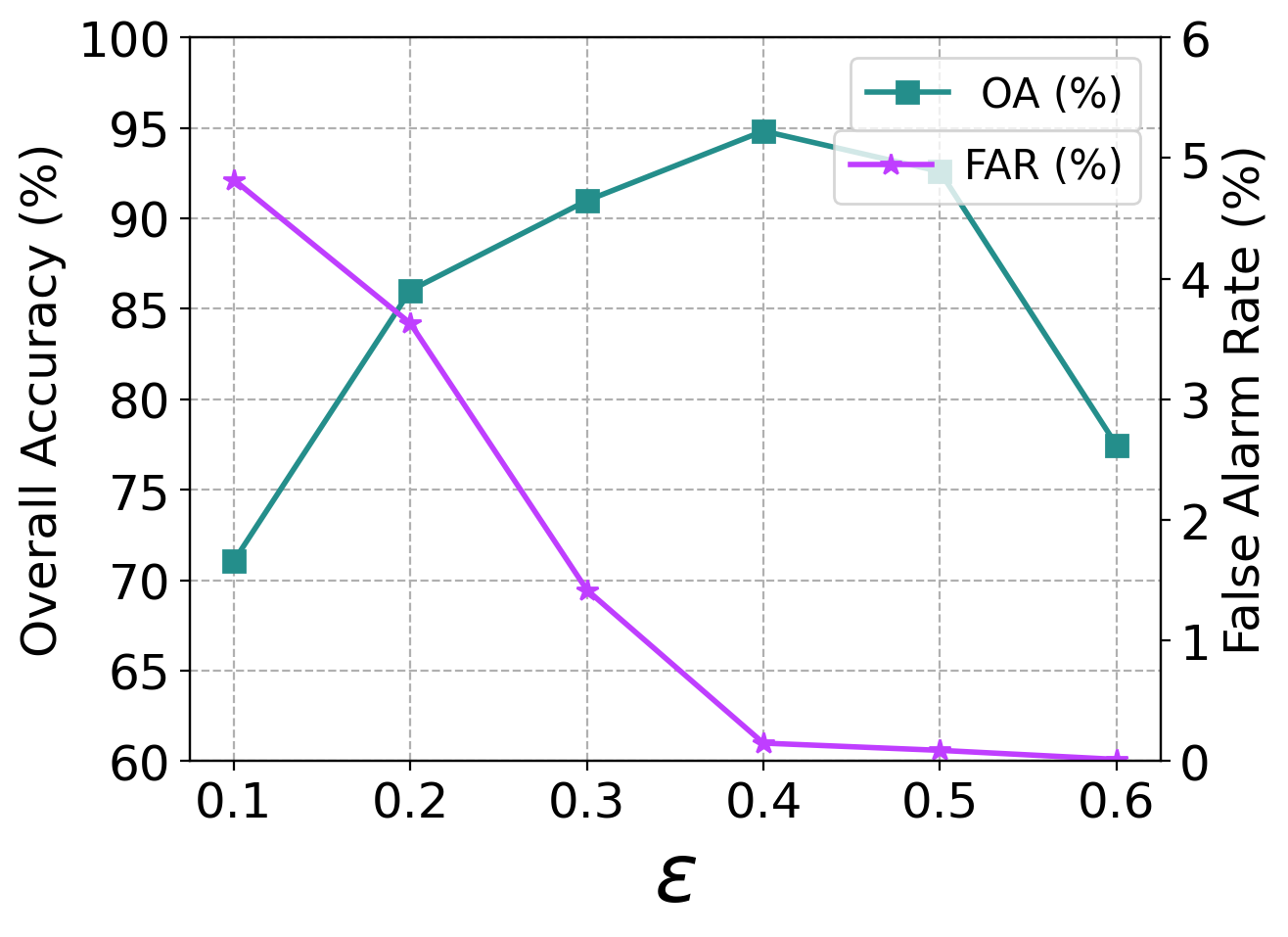}} \subfloat[$\gamma = 0.7$]{\includegraphics[width=4.5cm]{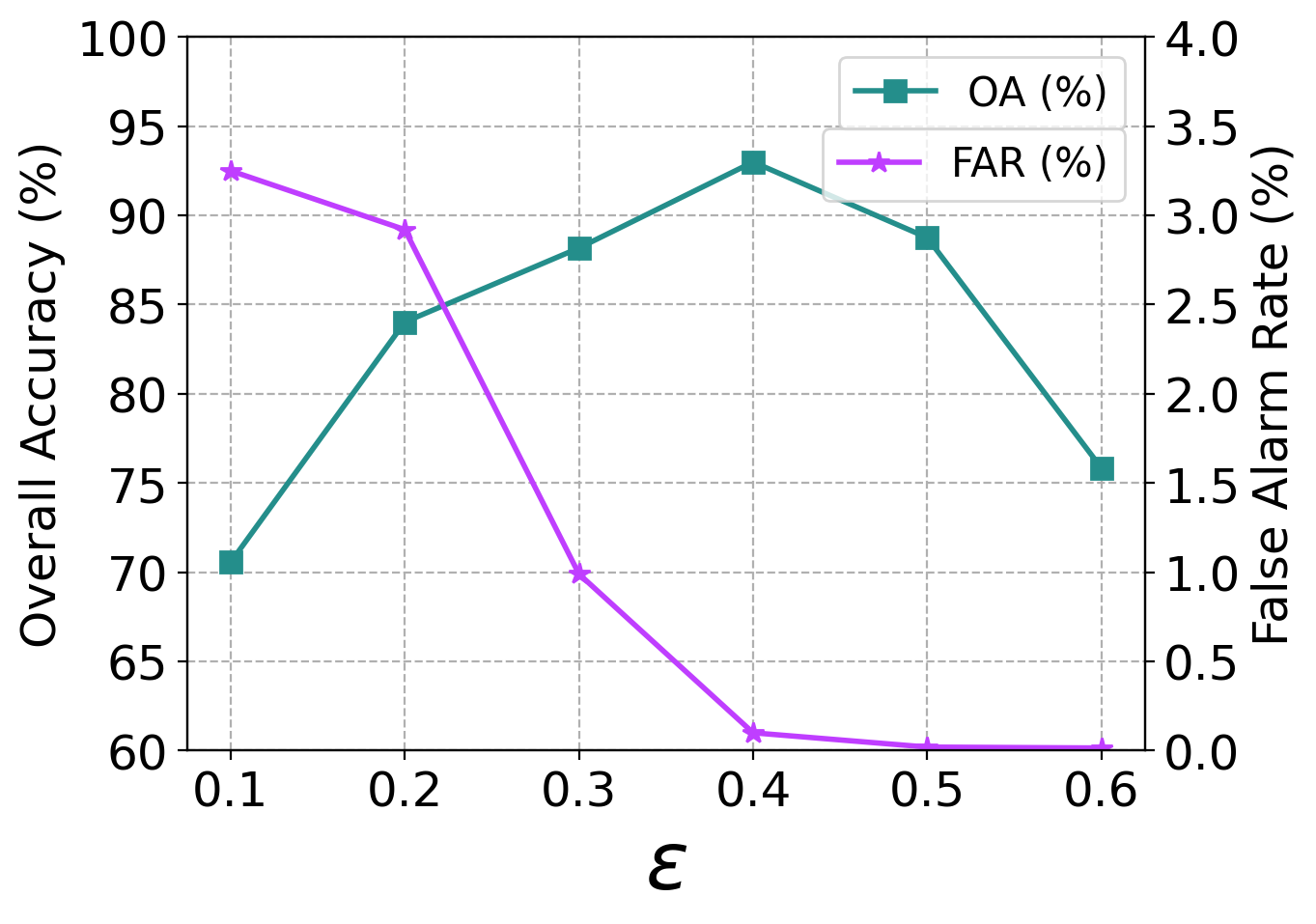}} \subfloat[$\gamma = 0.8$]{\includegraphics[width=4.5cm]{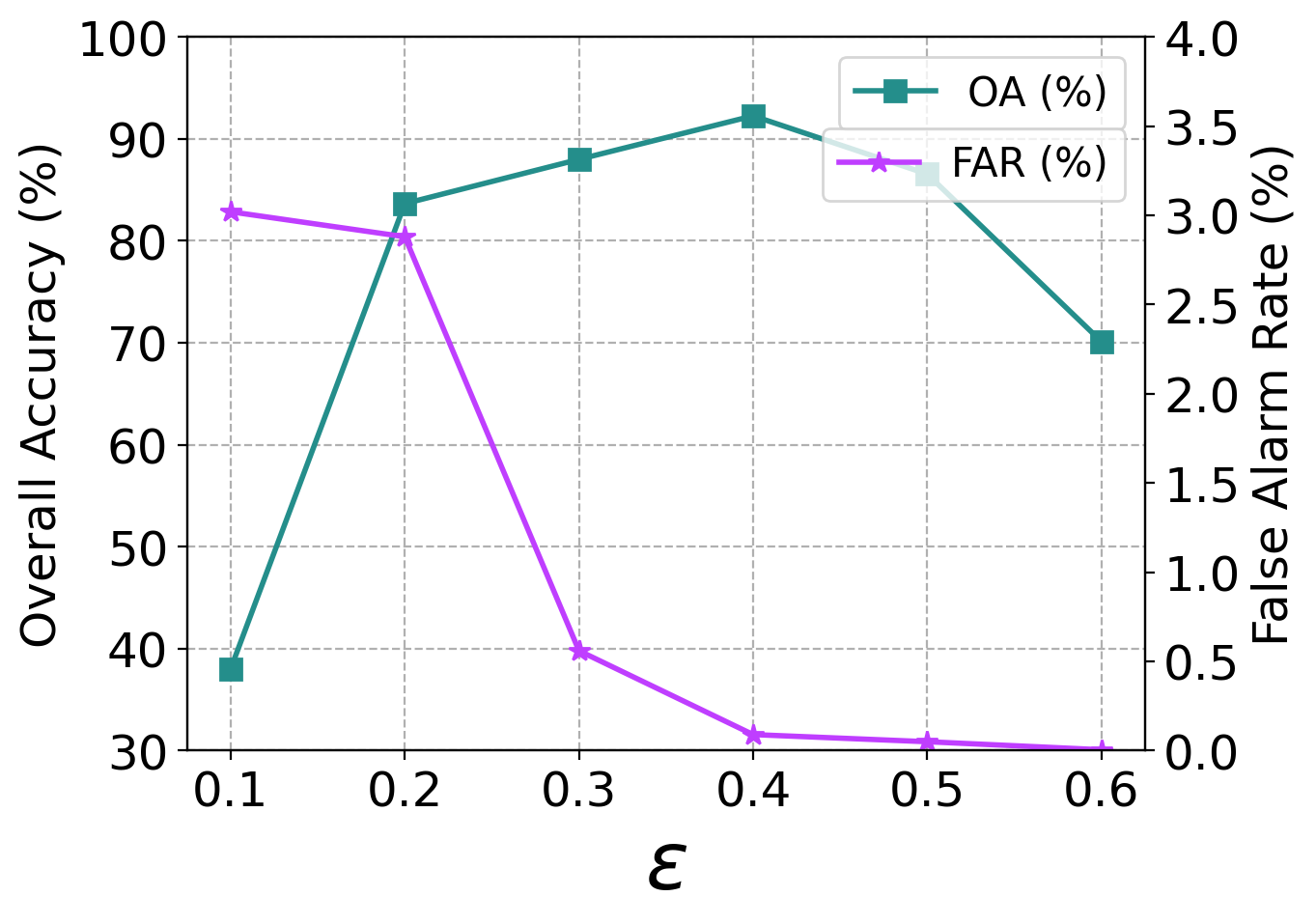}}

	\caption{Evaluations with different settings of detection threshold $\epsilon$ and coverage threshold $\gamma$. Magnifier is initialized with ScenarioA and validated with ScenarioC and ScenarioD (i.e., $S_A$ / $S_C$ + $S_D$). Besides, the detection duration is set to 15s (i.e., $\tau = 15 (s)$).}
	\label{fig_abl2}
\end{figure*}

We visualize the detection threshold $\epsilon$ and coverage rate $\gamma$ are discussed in Figure~\ref{fig_abl2}. The value of $\epsilon$ brings different trade-off between DR and FAR. Higher $\epsilon$ brings better FAR but poorer DR and vice versa. It shows that Magnifier achieves the best performance when $\epsilon = 0.4$ and $\gamma = 0.5$. Too higher of $\gamma$ will suffer from false negative while the opposite may bring out too much false alarms.

\section{Discussion}

\begin{table}[!t]

	\caption{Network access detection in multi-device scenario. DR with multi-labels is presented.} 
	\label{fig_dis1}
	\centering   
	
	\begin{tabular}{lcccc}  
		\toprule
		&\multicolumn{2}{c}{Brand Level}&\multicolumn{2}{c}{Model Level}\cr
		\cmidrule(r){2-3} \cmidrule(r){4-5} 
        &{\bfseries DR (\%)}&{\bfseries FAR (\%)}&{\bfseries DR (\%)}&{\bfseries FAR (\%)}\cr
		\midrule
  Sivanathan~\cite{sivanathan2018classifying}  & 82.73 & 15.85 & 74.15 & 18.89 \cr
        ProfilIoT~\cite{meidan2017profiliot} & 86.33 & 9.80 & 82.17 & 11.63 \cr
        \textbf{Magnifier}& {\bfseries 97.54} & {\bfseries 0.25} & {\bfseries 96.83} & {\bfseries 0.25} \cr

		\bottomrule
		
	\end{tabular} 

\end{table}
 
\subsection{Universality of Magnifier}
Commercial solutions employed by enterprises predominantly center around monitoring the network access of laptops or desktops, primarily utilizing endpoint-based methods. However, as highlighted in the paper, there is a notable scarcity of solutions claiming the capability to detect the network access of mobile devices, primarily due to the limitations inherent in such methods.

The inclusion of commercial solutions in our paper's discussion is a valuable addition. Moreover, we have augmented our experiments with additional baselines, specifically concentrating on IoT device recognition and encrypted traffic classification.

Sivanathan~\cite{sivanathan2018classifying} introduced a machine learning framework for the classification of IoT devices. This classification method involved analyzing various traffic features, including activity cycles, signaling patterns, communication protocols, and cipher suites.

ProfilIoT~\cite{meidan2017profiliot} represents a machine learning-driven approach designed for the identification of IoT devices within a network. The primary goal is to discern whether a given traffic flow originates from a PC, a smartphone, or a specific IoT device. The methodology involves the deployment of a set of classifiers, utilizing machine learning algorithms in a multi-stage process. This process enables the classification of traffic flows generated by a particular device based on their respective IP addresses.

We assessed the supplementary baselines under the identical settings outlined in the paper, specifically conducting the classification within the established detection duration of 15 seconds. The experimental results are illustrated in Table~\ref{fig_dis1}.

\subsection{Theoretical Analysis}
We have delved deeper into the efficiency of the distillation algorithm by providing a theoretical analysis of confidence scores.

The initial dnForest captures fingerprints from all traffic generated when devices connect to the network, encompassing both specific and background traffic. The presence of background traffic can result in significant false alarms during classification, given that these applications are commonly found on various devices. Moreover, background traffic constitutes a substantial portion of the captured data, causing an imbalance among dnTrees within a dnForest.

To address this challenge, we introduce a two-stage distillation algorithm designed to emphasize proprietary fingerprints while diminishing the importance of background ones. To further validate the efficiency of the algorithm, we conduct a comparison of confidence score distributions before and after fingerprint distillation. Specifically, for each test sample, we compute the confidence score alongside its corresponding fingerprint:

\begin{equation}
\begin{aligned}
{\rm Conf}^c_i = F(x_i^y,{\rm dnTree}^y)
\end{aligned}
\end{equation}

In the context of Magnifier, where $F(\cdot)$ represents the detector function, $x_i^y$ denotes a test sample from class y, and ${\rm dnTree}^y$ represents the fingerprint of class y. By averaging the scores of all test samples, we derive confidence scores that facilitate correct classification. Simultaneously, we compute the confidence score with respect to other fingerprints:

\begin{equation}
\begin{aligned}
{\rm Conf}^w_i = \frac{\sum\limits_{k\in Y,k \neq y} F(x_i^y,{\rm dnTree}^k)}{|Y|}
\end{aligned}
\end{equation}
where $|Y|$ is the number of the generated fingerprints. By averaging the scores of all test samples, we can obtain confidence scores indicative of incorrect classifications.

Before fingerprint distillation, the average ${\rm Conf}^c$ and ${\rm Conf}^w$ is 0.678 and 0.461, respectively. After the two-stage distillation, the average ${\rm Conf}^c$ and ${Conf}^w$ is 0.883 (30.2\% improvement) and 0.167 (63.8\% decline), respectively. The enhanced divergence of the two scores theoretically explain the experimental results in Section 6.5. 

Prior to fingerprint distillation, the average ${\rm Conf}^c$ and ${\rm Conf}^w$ are 0.678 and 0.461, respectively. Following the two-stage distillation, the averages for ${\rm Conf}^c$ and ${\rm Conf}^w$ become 0.883 (reflecting a 30.2\% improvement) and 0.167 (representing a 63.8\% decline), respectively. The amplified divergence between these two scores theoretically explains the experimental results outlined in Section~\ref{section_abl}.

\subsection{Practical Deployment}
Before deployment, the initialization of Magnifier involves the meticulous construction and distillation of fingerprints, as elaborated in Section~\ref{sec_method}. This process ensures the creation of well-defined fingerprints that form the basis for Magnifier's functionality. Subsequently, Magnifier has been effectively deployed in real-world scenarios. Specifically, it is employed for the detection of network access by mobile devices in both campus networks and production networks. The implementation is outlined as follows:
 
\begin{enumerate}

\item Magnifier is strategically deployed on both the gateway of our campus network and the enterprise network gateway. The implementation utilizes the KAFKA\footnote{Apache Kafka is an open-source distributed event streaming platform used by thousands of companies for high-performance data pipelines, streaming analytics, data integration, and mission-critical applications, which can be found at https://kafka.apache.org } and Protobuf\footnote{Protocol Buffers are language-neutral, platform-neutral extensible mechanisms for serializing structured data, which can be found at https://protobuf.dev} framework, supported by the DPDK\footnote{DPDK is the Data Plane Development Kit that consists of libraries to accelerate packet processing workloads running on a wide variety of CPU architectures, which can be found at https://www.dpdk.org} backend.

\item Initially, a probe, developed using DPDK, is employed to capture passing traffic at the gateway. Upon capturing DNS traffic, the subsequent network traffic within a 15-second timeframe from the identified IP is collected and stored in the cache. 

\item Subsequently, a feature extractor is utilized to consistently extract domain features from the cached traffic fragments. These extracted features are represented as "Dn-freq" pairs (e.g., {${dn}_1$: $n_1$, ${dn}_2$: $n_2$,..., ${dn}_k$: $n_k$ }) and encapsulated within a protobuf message. The KAFKA producer then pushes this message to a predefined topic. It's important to note that the processed traffic is promptly flushed from the cache.

\item The real-time detection algorithm is integrated into a Docker environment, allowing for efficient resource allocation by adjusting the number of detector instances. These instances continually retrieve protobuf messages from the subscribed topic. The algorithm utilizes well-designed fingerprints to detect behaviors and outputs results, including the determination of network access and identification of the device's brand/model.
 
\end{enumerate}

Magnifier was initially deployed on the gateway of our laboratory network. To assess the False Alarm Rate (FAR), we conducted an audit of the background network for 2 hours, during which no attempts of network access were made by mobile devices. Among the 20 thousand audited fragments, Magnifier triggered 37 alarms, resulting in a FAR of approximately 0.19\%. Subsequently, we evaluated the Detection Rate (DR) by simulating repeated connections of mobile devices to an internal WIFI. Magnifier achieved an impressive DR of 96.86\%, demonstrating performance consistent with our experimental results. Extending our deployment to an industrial enterprise, characterized by a high level of confidentiality, Magnifier maintained robust performance. In this context, it achieved a DR of 97.15\% and an exceptionally low FAR of 0.097\%.

\subsection{Privacy Issues}
Generating fingerprints for mobile devices using Magnifier does not inherently pose security and privacy issues, particularly in terms of user privacy disclosure. On one hand, Magnifier exclusively focuses on domain features within network traffic during fingerprint construction. Device-specific details, including serial numbers, MAC addresses, or IP addresses, are intentionally excluded from the fingerprint generation process. This deliberate omission of sensitive information ensures that even if the fingerprints were to be made publicly available, there would be no risk of exposing personally identifiable data or any device-specific identifiers. On the other hand, the publicly available dataset (NetCess2023) will not lead to privacy disclosure neither. Because most of the communication traffic is encrypted by TLS/SSL or private encryption protocol over HTTP. That means the payload between devices and servers is encrypted and invisible. The domain features Magnifier extracted are mainly from the headers of the encrypted traffic, which will not lead to the exposure of the communication detail.

\subsection{Potential Countermeasures}\label{counter_abl}

\quad\textbf{Concept of Drift. }To assess the stability of Magnifier, we conducted a system upgrade, including the latest versions of the operating systems and all possible pre-installed applications and components (e.g., OS, app markets, or music players). We then used Magnifier to detect the network access of up-to-date mobile devices using the original fingerprints. Under the optimal settings ($\epsilon = 0.4$, $\gamma = 0.5$, and $\tau = 15s$), Magnifier demonstrated stable performance, achieving a 95.29\% Detection Rate (DR), only dropping by 1.87\%. For instance, when we updated the WeatherAPP (from Ver14.3.302 to Ver14.4.300) and HarmonyOS (from Ver3.0.132 to Ver3.0.168) on a Huawei P30, the DR decreased from 99.08\% to 98.53\%. This demonstrates that system or proprietary application upgrades have minimal impact on Magnifier's performance. This is attributed to the fact that these upgrades rarely modify the original domain features, and any domain feature drift has minimal influence on the global fingerprints.

\textbf{Limitations. }While Magnifier exhibits effectiveness and efficiency in detecting the network access of mobile devices with prior knowledge, it encounters limitations when tasked with identifying novel classes. Nevertheless, network access patterns of novel classes often share similarities with known classes. Consequently, a template can be abstracted to fingerprint the network access of these novel classes.

\section{Conclusion} \label{sec_conclusion}
In this paper, we propose Magnifier, a method for detecting mobile device network access behavior. Magnifier utilizes lightweight fingerprints, constructed by forming a Domain Name Forest for each device. Enhanced by a two-stage distillation algorithm, Magnifier optimizes the initial fingerprinting by adjusting the importance of the Domain Name Trees within each Forest. Our experiments in two real-world scenarios demonstrate the effectiveness of our approach in efficiently classifying both initial and repetitive network access of various mobile devices in real time, at both the brand and model levels. Additionally, we have thoughtfully curated the NetCess2023 dataset by capturing network access traffic from a comprehensive range of mainstream mobile devices. NetCess2023 includes 10GB of traffic from 26 models across 7 brands and is now publicly accessible. We believe that this newly released dataset will significantly benefit real-world network applications.

\section{Acknowledgment}
This work was supported by the National Natural Science Foundation of China (NSFC) No.62376265.

\bibliographystyle{IEEEtran}

\bibliography{Magnifier}

\end{document}